\documentclass[11pt,a4paper]{article}

\usepackage[utf8]{inputenc}
\usepackage[left=2.75cm, right=2.75cm, top=3cm, bottom=3cm]{geometry}
\usepackage[english]{babel}
\usepackage{natbib}
\usepackage{authblk}
\usepackage{amssymb,amsfonts,amsthm,latexsym,mathtools}
\usepackage{graphicx,xcolor,pdflscape}
\usepackage{hyperref,url}
\usepackage{longtable}

\usepackage{tabu, makecell, array, colortbl, booktabs, multirow}
	\newcolumntype{C}[1]{>{\centering\arraybackslash}p{#1}}

\usepackage[ruled,vlined,linesnumbered]{algorithm2e}

\usepackage{tikz, pgfplots}
\usetikzlibrary{arrows,chains,matrix,positioning,scopes}
\makeatletter
\tikzset{join/.code=\tikzset{after node path={%
\ifx\tikzchainprevious\pgfutil@empty\else(\tikzchainprevious)%
edge[every join]#1(\tikzchaincurrent)\fi}}}
\makeatother
\tikzset{>=stealth',every on chain/.append style={join},
         every join/.style={->}}
\tikzstyle{labeled}=[execute at begin node=$\scriptstyle,
   execute at end node=$]

\makeatletter
\def\clineThicknessColor#1#2#3{\@ClineThicknessColor#1\@nil{#2}{#3}}
\def\@ClineThicknessColor#1-#2\@nil#3#4{%
	\omit
	\@multicnt#1%
	\advance\@multispan\m@ne
	\ifnum\@multicnt=\@ne\@firstofone{&\omit}\fi
	\@multicnt#2%
	\advance\@multicnt-#1%
	\advance\@multispan\@ne
	{\color{#4}%
		\leaders\hrule\@height#3\hfill}%
	\cr}
\makeatother

\newcommand{\N}{\mathbb N}
\newcommand{\R}{\mathbb R}
\newcommand{\calE}{\mathcal E} 
\newcommand{\calC}{\mathcal C} 
\newcommand{\calP}{\mathcal P} 
\newcommand{\bp}{\mathbf{p}}

\newcommand{\pr}{\mathbb{P}}
\newcommand{\esp}{\mathbb{E}}

\newcommand{\vol}{\mathrm{Vol}}
\newcommand{\diag}{\mathrm{diag}}
\newcommand{\cut}{\mathrm{cut}}

\newcommand{\cl}{\textrm{clique}}
\newcommand{\wcl}{\textrm{w-clique}}
\newcommand{\init}{\textrm{init}}
\newcommand{\aon}{\textrm{aon}}

\newcommand{\Ku}{Q^{\textrm{w-clique}}}
\newcommand{\Ka}{Q^{\textrm{strict}}}
\newcommand{\Kawdc}{Q^{\textrm{wsc}}}
\newcommand{\Chsymw}{Q^{\textrm{sym}}}
\newcommand{\Chsym}{\widehat{Q}^{\textrm{sym}}}
\newcommand{\ChAON}{Q^{\textrm{aon}}}

\DeclareGraphicsExtensions{.pdf, .jpg, .jpeg, .png, .gif,.eps}
\graphicspath{{figs/}{.}}

\title{Comparison of modularity-based approaches for nodes clustering in  hypergraphs}

\author[1]{Veronica Poda}
\author[2]{Catherine Matias}

\affil[1]{ University of Trento, Via Sommarive, 14, 38123, Povo, Italy. Email: \href{mailto:very.poda@gmail.com}{\texttt{very.poda@gmail.com}}}
\affil[2]{ Sorbonne Université, Université de Paris Cité, Centre National de la Recherche Scientifique, Laboratoire de Probabilités, Statistique et Modélisation, 4 place Jussieu, 75252 PARIS Cedex 05, France. Email: \href{mailto:catherine.matias@math.cnrs.fr}{\texttt{catherine.matias@math.cnrs.fr}}}

\date{}

\begin{document}

	\maketitle

\begin{abstract}
Statistical analysis and node clustering in hypergraphs constitute an emerging topic suffering from a lack of standardization. In contrast to the case of graphs, the concept of nodes' community in hypergraphs is not unique and encompasses various distinct situations.
In this work, we conducted a comparative analysis of the performance of modularity-based methods for clustering nodes in binary hypergraphs.  

To address this, we begin by presenting, within a unified framework, the various hypergraph modularity criteria proposed in the literature, emphasizing their differences and respective focuses. Subsequently, we provide an overview of the state-of-the-art codes available to maximize hypergraph modularities for detecting node communities in  hypergraphs. Through exploration of various simulation settings with controlled ground truth clustering, we offer a comparison of these methods using different quality measures, including true clustering recovery, running time, (local) maximization of the objective, and the number of clusters detected.

Our contribution marks the first attempt to clarify the advantages and drawbacks of these newly available methods. This effort lays the foundation for a better understanding of the primary objectives of modularity-based node clustering methods for binary  hypergraphs.\\

\textbf{Keywords}: community detection; higher-order interaction; hypergraph; modularity; node clustering
\end{abstract}

\section{Introduction}\label{sec:intro}

The interest in higher-order interactions stems from the recognition that many phenomena are inherently more complex than what can be effectively represented by pairwise relationships alone.
While graphs model pairwise interactions, hypergraphs generalize this concept by capturing higher-order interactions involving more than two elements. This extension provides a more expressive framework for modeling intricate dependencies and interactions in various fields, ranging from social network analysis \citep[early acknowledged in][]{simm:50} or co-authorship relations~\citep{roy:ravi:15} to ecological systems \citep{muyi:20}, neurosciences \citep{chela:21} or even chemistry \citep{flam:etal15}. 
We refer to \cite{batt:etal:20,bick:etal:21,torr:etal:21} for recent reviews on higher-order interactions.

\begin{figure}[htbp]
\centering
\includegraphics[width=\textwidth]{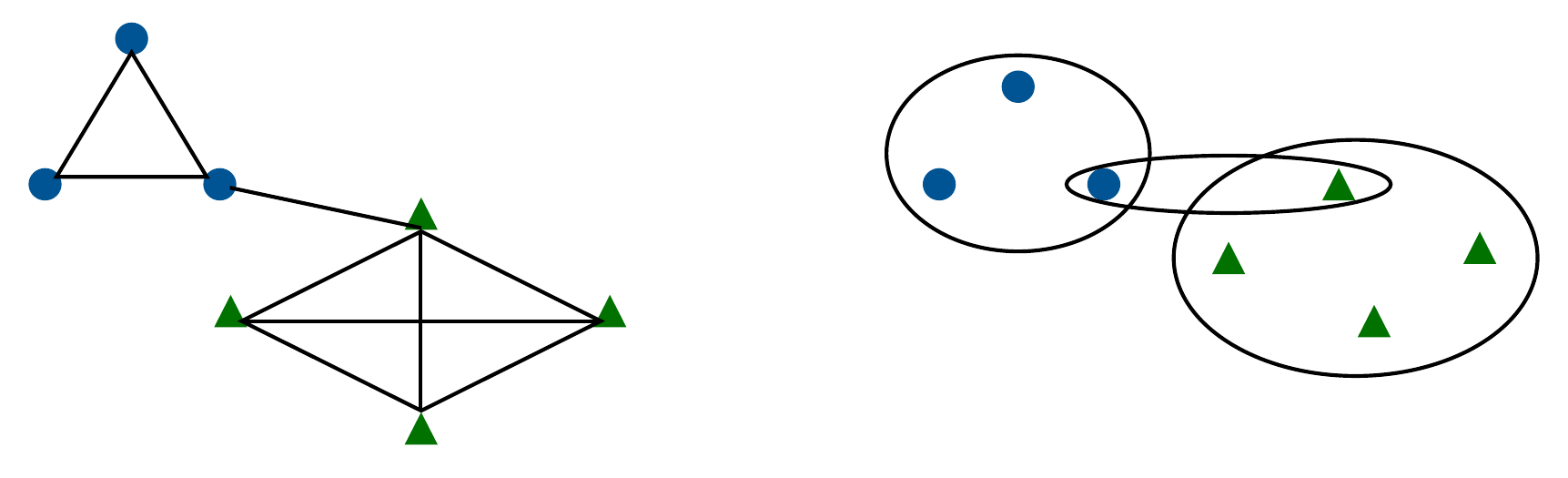}
\caption{On the left, a modular graph with two clusters is depicted, represented as circle-blue and triangle-green nodes, respectively. In each cluster, the number of within-cluster interactions is much larger than the between-clusters ones. On the right, a hypergraph is shown using the same set of nodes, where each clique from the previous graph is replaced by a hyperedge. In this hypergraph, the number of within-cluster interactions in each of the two clusters is the same as the number of between-clusters interactions. Is this hypergraph modular? Should we consider weighting hyperedges with respect to their sizes to analyze how modular the hypergraph is?}
\label{fig:hypergraph}
\end{figure}

With the emergence of hypergraph datasets \citep[see for e.g.][]{hg_char} to model higher-order interactions, the question of nodes clustering and, more specifically, the detection of communities in hypergraphs arises. In the context of graphs, the seminal paper by \cite{newm:girv:04} introduced the concept of modularity (commonly known as the Newman-Girvan modularity), paving the way for a flourishing literature on community detection in networks.
In the context of hypergraphs, the past few years have witnessed the surge of modularity-based proposals for hypergraph community detection. One of the first challenges is to propose a modularity criterion that measures the extent to which a hypergraph is composed of communities. This raises a more fundamental question: What is a community of nodes in a hypergraph? While in the context of graphs, a community is simply a set of nodes with more within-cluster interactions than between-clusters ones, generalizing that concept to hypergraphs is not immediate. As hypergraph interactions have a heterogeneous size (i.e., the number of nodes they contain), a primary issue is whether one should weigh the links with respect to (wrt) their sizes and put more emphasis on larger hyperedges (see Figure~\ref{fig:hypergraph} for an illustration). Consequently, various modularity criteria have recently emerged in the literature.

For a long time now, the computer science literature has tackled hypergraphs by simplifying them into graphs, employing two primary methods: the clique reduction graph, also known as the two-section graph, and the star-expansion graph. In the clique reduction graph, each hyperedge of a hypergraph is transformed into a clique in a graph over the same set of nodes (as illustrated in Figure~\ref{fig:hypergraph}, where the graph on the left represents the clique reduction of the hypergraph on the right). Conversely, the star-expansion graph constructs a bipartite graph by treating the original vertex set as the first part and introducing a new vertex for every original hyperedge in a second part. These parts are then connected whenever a node is contained in a hyperedge in the hypergraph.
While the former reduction loses information (the original hypergraph cannot be reconstructed from its clique reduction graph), the latter transformation is one-to-one,  given that the two parts are labeled (allowing the distinction between original nodes and original hyperedges) and hypergraphs with self-loops and multiple hyperedges are allowed.  
Consequently, a natural approach is to define hypergraph modularity by relying on graphs. Figure~\ref{fig:hypergraph} illustrates the limitations of such a method, where the clique reduction graph (on the left) appears clearly modular, while one may question whether the original hypergraph (on the right) should be considered modular or not.

In this article, we explore the current state-of-the-art and  challenges posed by modularity-based community detection methods in binary hypergraphs. In the context of graphs, \cite{Yang_etal_16} propose a comparative analysis of community detection algorithms for undirected and binary graphs. In the same vein, we here restrict our attention to modularity-based methods whose performances for community detection in binary hypergraphs are compared. 
The methodology is described in Section~\ref{sec:mat_met}. After introducing general notation, we first present a reformulated version of the different hypergraph modularities existing in the literature (Section~\ref{sec:mod}). The goal of this reformulation is to facilitate the comparison of concepts introduced independently from each other and never fully connected before. To be a valuable concept, a hypergraph modularity should come with a (local and/or heuristic) maximization algorithm that outputs a node clustering. Available implementations of such algorithms are presented in Section~\ref{sec:algos}. To compare the different modularities and maximization algorithms, it is mandatory to work with synthetic datasets where ground truth clustering is known and hypergraph statistics can be controlled. 
While in the graph context, recent years have seen the emergence of benchmark datasets for such a task, as for instance the  Lancichinetti-Fortunato-Radicchi benchmark graph \citep[LFR, ][]{LFR} used in \cite{Yang_etal_16}, there is  yet no  such benchmark for hypergraphs.  
We thus rely on several models for generating synthetic modular hypergraphs, described in Section~\ref{sec:synthetic}. Then Section~\ref{sec:experiments}  describes our experiments: which scenarios have been explored in each model generating method (Section~\ref{sec:scenario}) and quality assessment through the lens of different measures, namely true clustering recovery, running time, (local) maximization of the objective and the number of clusters detected (Section~\ref{sec:quality}).
All the results are presented in Section~\ref{sec:results} and a discussion   follows in Section~\ref{sec:discu}.  
The scripts to reproduce the experiments are available online (see details and links in Section~\ref{sec:available}).

To conclude this introduction, we mention that there are other methods to cluster the nodes of a hypergraph, such as spectral clustering approaches \citep{ghos:dukk:17:aos,chod:etal:siam:23} or model-based methods \citep{HSBM,ruggeri:etal:23}. 
It is also possible to cluster hyperedges instead of nodes \citep{ng:murp:22}. However, our focus in this work is on clustering nodes through modularity-based methods.

\section{Material and methods}
\label{sec:mat_met}
\subsection{General notation and definitions}

A hypergraph $H=(V, \calE)$ is  defined as a set of nodes $V=\{1,\dots, n\}$ and a set of hyperedges $\calE \subset \calP(V)$, where $\calP(V)$ is the set of all subsets of $V$. In other words,  each hyperegde $e\in \calE$ is a subset of nodes in $V$ (namely, $e\subset V$ or $e\in \calP(V)$). 
A hypergraph can either be binary (presence/absence of subsets of nodes) or  \emph{weighted} (also equivalently called \emph{multiple}). In the latter case, the hypergraph $H=(V, \calE, w)$ comes with a weight function $w:\calP(V) \to \N\cup \{0\}$ such that $\forall e \notin \calE$, we have  $w(e)=0$, and $w(e)\in \N$ otherwise. 
The weight counts how many times a hyperedge appears in the hypergraph. Multiple (i.e., weighted) hypergraphs can be viewed as hypergraphs where the set of hyperedges $\calE$ is allowed to be a multiset (some hyperedges may appear several times). 
A binary hypergraph is a particular case of a weighted hypergraph with weight function  being the indicator function $w(e)=1\{ v \in e\}$ (i.e., each hyperedge has multiplicity 1).
The size of a hyperedge $e$ is the number of nodes it contains $|e| =\sum_{v\in V} 1\{ v \in e\}$.  
A hypergraph is said to be \emph{$s$-uniform} if it only contains hyperegdes of size $s$. 
Any 2-uniform hypergraph is simply a graph. We let  $\calE_s$ denote the subset of $\calE$ of hyperedges with size $s$.
We can allow hyperedges $e\in \calE$ to be multisets of $V$, in which case nodes may appear more than once in the same hyperedge. Such hypergraphs are called \emph{multiset} hypergraphs and can  either be binary or multiple. 
In a multiset hypergraph, each node $v\in V$ has a multiplicity in hyperedge $e\in \calE$, denoted by $m_e(v) \in \N\cup \{0\}$,  which counts the number of times this node appears in that hyperedge.  
Moreover, the hyperedge size accounts for the nodes multiplicity and becomes $|e|=\sum_{v\in V} m_e(v)$. For example, a self-loop $\{u,u\}$ is a (multiset) hyperedge of size 2. 
In the following, unless otherwise stated, all sets can be multisets in which case all counts include multiplicities (be it for nodes or for hyperedges). A  hypergraph is  said \emph{simple} whenever it is  binary and non-multiset, i.e.,  neither nodes or hyperedges may be repeated. 
The (weighted) degree $\deg_H(v)$ of a node $v$ in a hypergraph $H$ is  the (weighted) count of the hyperedges it belongs to, 
namely $\deg_H(v)=\sum_{e\in \calE}  w(e)$.
The incidence matrix  $H$ of the hypergraph has  size $|V|\times |\calE|$  and entries $H(v,e)=1\{ v \in e\}$ or $m_e(v)$ for multiset hypergraphs. 
Note that we use the same notation $H$ for a graph and its incidence matrix, the difference should be clear from the context. Letting $w=(w(e))_{e\in \calE}$ denote the (column) vector of the hyperedges weights and  $w^\intercal$ its  transpose, we obtain the vectors of node degrees and hyperedges sizes as $Hw$ and $w^\intercal H$, respectively. 
Two nodes are said \emph{incident} whenever they belong to a same hyperedge $e\in \calE$.

For any subset of nodes $C\subset V$, we define its volume:  
\[
\vol_H(C) = \sum_{v\in C} \deg_H(v) , 
\]
and the (weighted) number of hyperedges whose nodes are all included in $C$: 
\[
 e_H(C) = \sum_{e\subset C} w(e).
 \]
Note that $\vol_H(V) =\sum_s s |\calE_s|$ and $e_H(V)= |\calE|=\sum_s  |\calE_s|$.

From a (weighted) hypergraph $H=(V, \calE)$ we may construct its clique reduction graph $G^{\cl}=(V,E)$. This graph has the same set of nodes $V$ as the hypergraph and every hyperedge $e\in \calE$ in the hypergraph is \emph{reduced} into a complete clique in the graph. In other words, for any hyperedge $e\in \calE$ with size $|e|\ge 2$ and for any pair of incident nodes $u, v\in e$, the graph $G^{\cl}$ contains the edge $\{u,v\}\in E$ and only edges obtained in this way are contained in $E$. The (weighted) adjacency matrix $A^{\cl}$ of the clique reduction graph satisfies $A^{\cl}=H \diag(w)H^\intercal $, where $H$ is the incidence matrix of the hypergraph 
and $\diag(w)$ is the diagonal matrix induced by the vector of hyperedge weights $w$. In general, self-loops are removed from $A^{\cl}$ and $A^{\cl}_{uu}$ is set to $0$ for any $u\in V$. This can be done directly by setting $A^{\cl}=H \diag(w)H^\intercal -D_V$ where $D_V$ is the diagonal matrix of node degrees $(\deg_H(v))_{v\in V}$.

A nodes clustering is a partition $\calC=(C_1,\dots, C_K)$ of the set of nodes $V$ into parts called clusters.  For any partition $\calC=(C_1,\dots, C_K)$ of the set of nodes $V$ and any subset $e\subset V$, we let $e \cap \calC=(e_1,\dots, e_J)$ denote the partition of the subset $e$ induced by $\calC$. It has $J$ parts with $J\le K$ and is indeed a partition of $e$, namely 
 \[
 \forall j \neq  k  \in \{1,\dots, J\}, \quad  e_j\neq \emptyset , \qquad e_j\cap e_k=\emptyset , \qquad \cup_{j=1}^J e_j =e.
 \]
 The \emph{adjacent} clusters of a node $u\in V$ are the parts $C_k$ that contain at least one node $v\in C_k$ that is incident to $u$, or in other words such that there is a hyperedge $e\in \calE$ such that $u,v\in e$.

In this manuscript, the identity matrix is denoted by $I$ (its size should be clear from the context).
We already used notation   $|S|$ for the cardinality of a set $S$ (or a multiset),  and $1\{\mathcal{S}\}$ for the indicator function of an event $\mathcal{S}$.

\subsection{Modularities in hypergraphs}
\label{sec:mod}
Different hypergraph modularity criteria have been proposed in the literature up to now \citep{kumar:etal:20,kami:etal:19,kami:etal:21,Chodrow_21}. We recall these different quantities, using a unified presentation that highlights similarities and differences between them. As we will see, these are all constructed in the same way, namely  the  difference between a first term that is a specific hyperedge count and a second term that in some cases  corresponds to the expected value of this count under some null model, and otherwise is a correction term.  The differences between the expressions of those hypergraph  modularities come from: i) the type of hyperedges that are counted; ii) the null model used for computing the expectation or the correcting term; iii) possible weights to each of these terms. \\

\cite{kumar:etal:20}'s definition of hypergraph modularity corresponds to a graph modularity  as originally defined in \cite{newm:girv:04} and applied to a specific graph choice. 
Considering the clique reduction graph of a hypergraph, \citeauthor{kumar:etal:20} noticed that the reduction does not preserve the node degrees: in the clique reduction graph $G^{\cl}$, the degree of a node differs from its  initial value in the hypergraph $H$. Indeed, a simple computation shows that
\[
\deg_{G^{\cl}}(v) = \sum_{e \in \calE} H(v,e) w(e)(|e|-1) \ge \sum_{e \in \calE} H(v,e) w(e)= \deg_H(v). 
\]
Thus, \cite{kumar:etal:20} simply modified the weights in the clique reduction graph to preserve these degrees. 
Let $D_{\calE}=\diag(|e|)_{e \in \calE}$  denote the diagonal matrix of the hyperedges sizes. We define the \emph{weighted clique reduction graph} $G^{\wcl}$ through its adjacency matrix $A^{\wcl} =(A_{uv}^{\wcl})_{u,v \in V}$ by 
\[
A^{\wcl} = H \diag(w) (D_{\calE}-I)^{-1}H^\intercal  -D_V.
\]
The node degrees in this graph $G^{\wcl}$ are  equal to the initial node degrees in the hypergraph $H$ (where self-loops are discarded). This construction is equivalent to saying that for each hyperedge $e\in \calE$, we create $G^{\wcl}$ by forming a total of $\binom{|e|}{2}$ edges with weights $w(e)/(|e|-1)$, between any pair of  nodes incident in the hypergraph $H$. 

Then for any hypergraph $H=(V,\calE)$ and any partition $\calC=(C_1,\dots,C_K)$ of its set of nodes $V$, we let 
\begin{align}
\Ku (H,\calC) &= \frac 1 {2|\calE|} \sum_{k=1}^K \sum_{u,v \in C_k} \left( A_{uv}^{\wcl} - \frac{\deg_H(u)\deg_H(v)}{\sum_{i=1}^n \deg_H(i)} \right) \nonumber \\
&= \frac 1 {2|\calE|} \sum_{k=1}^K  \left(  
e_{G^{\wcl}}(C_k) 
- \frac{(\vol_H(C_k))^2}{\vol_H(V)} \right).  
\label{eq:Kumar_mod}
\end{align}
Note that $\Ku$ ranges in $[-1;1]$. It is an average over all pairs of nodes $u,v$ belonging to the same cluster $C_k$ of the difference between the weighted edge value $A_{uv}^{\wcl}$ in the weighted clique reduction graph and its expectation under a configuration model \citep{Chung:Lu} that accounts only for nodes degrees and plays the role of a null model. 
A high value of modularity $\Ku$ means  dense connections in the weighted clique reduction graph $G^{\wcl}$ between the nodes within the same cluster and sparse connections between nodes in different clusters. Going back to the hypergraph $H$, that means node pairs $u,v\in V$ belonging to the same cluster participate more in the same hyperedge than node pairs in different clusters.  \\

\cite{kami:etal:19} introduce a \emph{strict} hypergraph modularity such that only the hyperedges $e\in \calE$ entirely included in a same cluster contribute to increasing modularity, which is in sharp contrast with the previous proposal.   
For any hypergraph $H=(V,\calE)$ and any partition $\calC=(C_1,\dots,C_K)$ of its set of nodes $V$, we let 
\begin{equation}
\label{eq:Kami_mod}
\Ka(H,\calC)= \frac 1 {|\calE|} \sum_{k=1}^K \left( e_H(C_k) - \sum_{s\ge 2} |\calE_s| \left( \frac{\vol_H(C_k)}{\vol_H(V)} \right)^s \right). 
\end{equation}
Note that $\Ka$ also ranges in $[-1;1]$. 
Here, the first term inside the sum accounts for the number of hyperedges whose all nodes are within the same cluster. The second term comes from a generalization of the \citeauthor{Chung:Lu} model to hypergraphs. Again, it plays the role of an expected value of the first term $e_H(C_k)$ under some null model which preserves both node degrees and the (weighted) number $|\calE_s| $ of size-$s$ hyperedges. This quantity is  called by its authors the \emph{degree tax}. 
\\

\cite{kami:etal:21} propose a more general modularity that accounts for the \emph{homogeneity} of each hyperedge, namely, the fraction of its vertices that belong to the largest cluster (provided it is more than 50\%). For any subset $C\subset V$, any size $s\ge 2$ and any integer $c\in \{\lfloor s/2\rfloor +1,\dots,s\}$, we let $e_H^{s,c}(C)$ denote the number of size-$s$ hyperedges that have exactly $c$ nodes included in their majority part $C$. With our previous notation, we have
\[
e_H(C) = \sum_{s \ge 2} e_H^{s,s}(C).
\]
In the following, $\pr(\textrm{Bin}(s, p)=c)= \binom{s}{c} p^c(1-p)^{s-c}$ is the probability that a Binomial random variable with parameters $(s, p)$ takes the value $c$. 
Then for any partition $\calC=(C_1,\dots, C_K)$ of the set of nodes, \cite{kami:etal:21} introduce the modularity 
\begin{equation}
\label{eq:Kami_mod_wdc}
\Kawdc(H,\calC)= \frac 1 {|\calE|} \sum_{k=1}^K \sum_{s\ge 2} \sum_{c=\lfloor s/2\rfloor +1}^s  w_{s,c} \left[ e_H^{s,c}(C_k) - |\calE_s| \pr \left(\textrm{Bin}\Big(s, \frac{\vol_H(C_k)}{\vol_H(V)}\Big) =c\right) \right], 
\end{equation}
where $w_{s,c}\in [0,1]$ are hyper-parameters to be specified. 
Note that we have 
\[
\pr \left(\textrm{Bin}\Big(s, \frac{\vol_H(C_k)}{\vol_H(V)}\Big) =s \right) =  \left( \frac{\vol_H(C_k)}{\vol_H(V)} \right)^s
\]
so that \eqref{eq:Kami_mod} is a special case of \eqref{eq:Kami_mod_wdc} where $w_{s,c}=1\{c=s\}$.
Different setups may be considered for  the hyper-parameters $w_{s,c}$ and we focus here on the choices for which an optimisation algorithm is available, namely
\begin{equation}
\label{eq:wsc}
w_{s,c}=\left\{
\begin{tabular}{ll}
$1\{c=s\}$ & \text{strict setting},\\
$1\{c>s/2\} $& \text{majority setting},\\
$c/ s 1\{c>s/2\}$ & \text{linear setting}.
\end{tabular}
\right. 
\end{equation}
As already mentioned, the \textbf{strict} setting  
gives back $\Ka$, already introduced in \eqref{eq:Kami_mod}. 
For the other 2 settings, we call the corresponding modularities  $Q^{\textrm{majority}}$ and  $Q^{\textrm{linear}}$, respectively. 
\\

Finally, \cite{Chodrow_21} first defined a general \emph{symmetric} modularity, where for any partition $\calC$ of the set of nodes, the contribution of a  hyperedge $e\in \calE$ to the modularity of this partition is characterized  only by the vector $\bp$ whose   entries $p_k$ count the number of nodes in $e$ belonging to the $k$-th largest part in $e\cap \calC$. 
It is based on a general affinity function $\Omega: \calP \to \R$ that modulates the weight of the contribution of each partition vector $\bp$, where the set of partition vectors is
\[
\calP = \{\bp=(p_1,\dots p_J) ; p_1\ge  \dots \ge p_J \ge 1, \text{ for some } J\ge 1\}.
\]
For instance, a $s$-tuple of nodes with $s=7$ that are clustered by a partition $\calC$ into the parts $\{v_1\}; \{v_2,v_3\}; \{v_4\}; \{v_5,v_6,v_7\}$ induces the partition vector $\bp=(3,2,1,1)$. The symmetric modularity from \cite{Chodrow_21} will thus account for the different clusters counts that compose a hyperedge, treating all the clusters in an exchangeable way. 
We present the details of this modularity in Section~\ref{sec:SM_Chsym} from the Appendix. 
Then, the authors consider particular cases of their general symmetric modularity, relying on specific forms of the affinity function $\Omega$ (see Table 1 in that reference). However, an implementation of the algorithm for optimising the induced specific modularities is available  only for the \emph{all-or-nothing} affinity function on which we focus now.

The all-or-nothing modularity function is  defined as: 
\begin{align}
\label{eq:ChodAON_mod}
\ChAON (H,\calC) 
&= \sum_{k=1}^K \sum_{s\ge 2} \hat \beta_s \left(  \sum_{C_k' \subset C_k; |C_k'|=s} e_H(C_k') - \hat \gamma_s (\vol_H(C_k))^s 
\right) , 
\end{align}
where $\hat \beta_s$ and $\hat \gamma_s$ are parameters estimated from the data. 
While in general we may expect that both $\hat \beta_s, \hat \gamma_s >0$ (see Section~\ref{sec:SM_AON} in the Appendix for more details on these parameters), we then recover in this expression a sum of difference terms between a count of specific hyperedges, namely those entirely included in a cluster, and a correcting volume term. 
The extra parameters $\hat \beta_s, \hat \gamma_s$ might not seem natural at first. In fact, they appear as the result of an approximate maximum likelihood approach in a specific degree-corrected hypergraph stochastic blockmodel (DCHSBM), in the same way as \cite{newm:16}  did in a graph context.

As a final remark, \cite{Chodrow_21}  notice that considering the specific choices $\hat \beta_s=1$ and $\hat \gamma_s = |\calE_s| / \vol_H(V)^s$ in their modularity $\ChAON$, they recover  (up to a scaling factor and an additional term not depending on the partition $\calC$ and which can thus be discarded) the expression of the modularity  $\Ka$ from \eqref{eq:Kami_mod}. However, they argue that leaving these parameters free (adapting to the data) lends important flexibility to their approach.

\paragraph*{Additional comments.}
We already highlighted  similarities and differences between the different modularities defined above. Let us add some more comments.

Two extreme cases are represented by the modularities $\Ku$ and $\Ka$, the former being less stringent than the latter. 
Whenever a hyperedge is split by the partition $\calC$ into different clusters, it will be ignored by $\Ka$ but as soon as this hyperedge contains at least 2 nodes in the same cluster, the modularity $\Ku$ will account for it. The weakness in  $\Ku$ lies in that the exact composition of each hyperedge in nodes falling into the different clusters is captured only through pairs of nodes. 
The modularity $\Kawdc$ represents a  compromise between the 2 previous extremes: it accounts for \emph{homogeneous} hyperedges, namely hyperedges such that (at least) half of their nodes  fall into  a cluster that becomes a majority cluster. In particular, \cite{kami:etal:21} argue that the hyper-parameters $w_{s,c}$ may be chosen so that $\Kawdc$ well approximates $\Ku$ because contributions in the latter from parts that contain at most $s/2$ vertices may often be neglected. 
Finally, the modularity $\ChAON$ is as strict as $\Ka$ and focuses on hyperedges with nodes split into a unique cluster by the partition $\calC$. As already stressed, the major difference between  $\Ka$ and $\ChAON$ lies in that the latter, while summing similar differences as the former, weights differently each terms in those differences (with weights adaptive to the data, as they are estimated from these). 

Note that possible self-loops in the hypergraph $H$ never contribute to a modularity and may thus be discarded from the dataset. 
However, we highlight that all these modularities are developed for multiset hypergraphs, where nodes may be repeated in a same hyperedge. In particular, the \citeauthor{Chung:Lu} null models (for graphs and hypergraphs) used in defining modularities $\Ku, \Ka$ and $\Kawdc$ as well as the  DCHSBM underlying  the definition of the  modularity  $\ChAON$, all 
rely on models for multiset hypergraphs. While it is known in the case of graphs that this is inadequate \citep{Massen_Doye05,Cafieri_2010,Squartini_2011}, that assumption has not yet been discussed in the context of hypergraph modularities. It might be that the computational simplifications enabled by this assumption  prevent from any attempt not to use it \citep[see for e.g. Section B2 in Supplementary Material from][]{HSBM}.

\subsection{Modularity maximization methods}
\label{sec:algos}
In this section, we focus on available implementations for hypergraph nodes clustering through modularity-based methods. We briefly describe the corresponding algorithms and their major characteristics, as  well as  the options that were chosen for our comparison study. All the algorithms require an initialization, most of the times relying on an initial partition where each node is in its own part, i.e., $\calC^{\text{own}}=(\{1\},\dots,\{n\})$. 
We group the different methods by the packages where they can be found. 
A summary is given in Table~\ref{tab:algos}. 

Note that we did not include in our experiments a comparison with methods based on clique reductions. Indeed, \cite{kumar:etal:20} already did so and concluded that ``hypergraph based methods perform consistently better than their clique based equivalents'' (end of page 16 in that reference).

\paragraph*{HyperNetX package.}
The \texttt{HyperNetX} Python package \citep{HyperNetX} contains a \texttt{modularity} submodule (see \url{https://pnnl.github.io/HyperNetX/modularity.html}) including various functions for hypergraph clustering through modularity maximization.

\cite{kumar:etal:20} propose to maximize their modularity $\Ku$ relying on the popular and fast Louvain algorithm for graphs \citep{Louvain}. More precisely, they do not simply apply Louvain algorithm on the graph $G^{\wcl}$ but rather propose an Iteratively Reweighted Modularity Maximization (\texttt{IRMM}) algorithm where they iteratively apply Louvain on a weighted clique reduction graph, and compute new hyperedge weights \cite[see Algorithm 1 in][]{kumar:etal:20}. The hyperedge re-weighting step puts a larger weight on hyperedges which are cut into more \emph{unbalanced} partition vectors  by the current partition $\calC$. For example, a size-$s$ hyperedge cut into the partition vector $\bp=(s-1,1)$ (meaning a unique node falls in a cluster different from the majority one) is much more unbalanced than another one cut into the partition vector $\bp=(s/2,s/2)$ (namely half of the nodes belong to a first cluster, the other half belonging to a second cluster) and thus gets a larger weight \citep[see Figure 1 in][]{kumar:etal:20}. By getting a larger weight, it is more likely that the unique node in this hyperedge will join the majority cluster at Louvain's next step.  The function \texttt{hmod.kumar} implements the \texttt{IRMM} algorithm. \\
 
The last step refinement (\texttt{LSR}) is an algorithm described in \cite{kami:etal:21}. This is a general method that starting from an initial partition of the nodes, iteratively moves one vertex at a time (in a random order) to a neighboring cluster whenever it improves $\Kawdc$, until convergence. The authors propose to start by running the \texttt{IRMM} on the weighted clique reduction graph, then the resulting partition is used as initialisation in their \texttt{LSR} procedure, that aims at maximizing $\Kawdc$. For the specific choices \texttt{strict, majority} and \texttt{linear} of the hyper-parameters $w_{s,c}$ described above, implementations are provided. The modularity $\Kawdc$ is obtained through the function \texttt{hmod.modularity} from  the \texttt{HyperNetX}  package and the \texttt{LSR} algorithm is implemented in the function \texttt{hmod.last\_step} from this same package. Both functions contain the 3 different options for hyper-parameters $w_{s,c}$ defined in \eqref{eq:wsc} and the default choice is \texttt{linear}. This is this option that we choose for our comparisons.

\paragraph*{strictModularity package.}
\cite{kami:etal:19} propose a Clauset-Newman-Moore like (\texttt{CNM-like}) algorithm to maximize $\Ka$ \citep[see][for the original CNM algorithm]{Claus:Newm:04}.  Starting with  partition $\calC^{\text{own}}$ where each node is in its own part, this algorithm iterates over the set of hyperedges that are split into more than 2 clusters by the current partition, trying to merge all the parts it touches and looking for a modularity improvement. 
More precisely, the algorithm comes in two versions. In the first one, a loop over all hyperedges is taken, so that at each step all hyperedges are searched and evaluated for merging. In the second one, a stochastic approach is taken which evaluates at each step just one randomly chosen hyperedge \citep[see Algorithm 1 in ][for more details]{kami:etal:19}.  The stochastic version  is computationally less expensive, especially for larger hypergraphs; however it requires to set a maximal number of iterations.   In what follows, we choose that second version and  set the number of iterations to twice the total number of hyperedges. 
The implementation is available from \cite{kami_code:19}, in a mix of Python and Julia files. More precisely, a script \texttt{strictModularity.py} contains a ``quick'' Python implementation that should work on small datasets only, while a Julia function \texttt{find\_comms} is more generally provided to perform the \texttt{CNM-like} algorithm. We rely on the latter in our experiments.

\paragraph*{HyperModularity  package.}
\cite{Chodrow_21} propose  the  Hypergraph Maximum Likelihood Louvain (\texttt{HMLL}) algorithm to maximize their symmetric  modularity (defined in Section~\ref{sec:SM_Chsym} from the Appendix) and  the  simpler and faster \texttt{AON-HMLL} algorithm for maximizing the specific all-or-nothing $\ChAON$ modularity. 
Both the \texttt{HMLL} and the \texttt{AON-HMLL} are implemented in the Julia package \texttt{HyperModularity} \citep{Hyper-modularity}.
However   the current version of the \texttt{HyperModularity} package does not contain an implementation of an estimation of a general affinity function $\widehat{\Omega}$ that is required to compute the symmetric modularity. That is why we focus on the AON modularity $\ChAON$ and the corresponding \texttt{AON-HMLL} algorithm.

The  \texttt{AON-HMLL} algorithm is an iterative algorithm that mimics the standard graph Louvain algorithm in that it starts with initial configuration $\calC^{\textrm{own}}$ (each node is in its own part) and at the first iteration, it greedily moves nodes to adjacent clusters (i.e., clusters that contain incident nodes) until no more improvement of $\ChAON$ is possible. The subsequent iterations however differ from Louvain's approach and instead of considering a weighted graph on ``supernodes'', it greedily moves entire clusters to adjacent ones whenever this improves  $\ChAON$. Note that the option \texttt{startclusters} from \texttt{Simple\_AON\_Louvain\_mod} determines which initial partition is used to estimate the parameters $\hat \beta_s, \hat \gamma_s$. We rely on \texttt{startclusters == "cliquelouvain"} that gives the best results in general. 
%

\begin{table}[htbp]
\centering
\begin{tabular}{C{3.9cm}C{1.6cm}C{1.7cm}C{1.7cm}p{4.3cm}}
\toprule
Function (package or script) & Modularity & Algorithm &  Language & Options choices\\
\hline
\texttt{hmod.kumar} 
\citep[\texttt{HyperNetx},][]{HyperNetX} 
&$\Ku$ &\texttt{IRMM}&  Python & Init: $\calC^{\text{own}}$\\
\hline
\texttt{hmod.last\_step}  
\citep[\texttt{HyperNetx},][]{HyperNetX}
& 
$Q^{\textrm{linear}}$& \texttt{LSR}& Python & Init: Output(\texttt{IRMM}), $w_{s,c}$= linear\\
\hline
\texttt{find\_comms} 
\citep{kami_code:19} 
&$\Ka$ &\texttt{CNM-like}&  Julia &  Init: $\calC^{\text{own}}$,  Stochastic version  \\ 
\hline
\texttt{Simple\_AON\_Louvain}
\citep[\texttt{HyperModularity},][]{Hyper-modularity}
&$\ChAON$& \texttt{AON-HMLL}& Julia&  Init: $\calC^{\text{own}}$, \texttt{startclusters == "cliquelouvain"} \\
\bottomrule
\end{tabular}
\caption{Summary of functions (with package name and reference) for clustering hypergraphs through modularity-based approaches. We indicate which modularity is maximized by the function (second column), the corresponding algorithm (third column), the implementation language (fourth column) and  our option choices (fifth column).}
\label{tab:algos}
\end{table}

\subsection{Synthetic models for binary and modular hypergraphs} 
\label{sec:synthetic}

To compare the different modularity-based approaches for clustering hypergraphs nodes, it is mandatory to rely on simulations of modular hypergraphs where ground truth clusters are known. 
As mentioned earlier, there is no single standard method for generating modular hypergraphs, and, to our knowledge, there are two main approaches. The first approach is based on hypergraph stochastic block models, with several variants proposed in the literature. The second approach involves a generalization of the LFR model for graphs \citep{LFR}. 
We chose to consider two variants of the first approach and the only one that we are aware of in the second approach. A  summary of these models is given in Table~\ref{tab:models}. 
We highlight the similarities and differences between those different generating models and the characteristics of the hypergraphs generated by those approaches. 

In all those models, we fix a number of nodes $n$, either fix or randomly generate a true number of clusters $K$ (that might depend on $n$) as well as a true partition of the nodes $\calC^{\text{true}}=(C_1^{\text{true}} ,\dots, C_K^{\text{true}} )$ and a maximal size $S$ of hyperedges.

\paragraph*{Hypergraphs with HSBM.}  
We consider datasets simulated under a simple (binary and non-multiset) Hypergraph Stochastic Blockmodel \citep[HSBM, see][]{HSBM} generated through the \texttt{R} package \texttt{HyperSBM} \citep{HyperSBM}. 
In this model, we fix the true number of clusters $K$, their proportions $\pi=(\pi_1,\dots, \pi_K)$ such that $\pi_k\in (0,1)$ and $\sum_k \pi_k=1$ and the following parameters, for any $2\le s \le S$, 
\begin{align*}
\alpha_{s}
&= \pr(e \in \calE_s | \exists 1\le k\le K,  e \subset C_k^{\text{true}} ) ,\\
\beta_{s}
&= \pr (e \in \calE_s | \forall 1\le k\le K,  e \not\subset C_k^{\text{true}} ), 
\end{align*} 
so that $\alpha_s$ (resp. $\beta_s$) is the probability for a $s$-tuple of nodes to form a hyperedge given that they belong to the same cluster (resp. given that they are not all in same cluster). 
The parameters should be chosen in order to ensure that the generated hypergraphs are  modular. 
To this aim, we consider the ratios $\rho_s$ of the number of within-cluster size-$s$ hyperedges over the number of between-clusters size-$s$ hyperedges obtained as: 
\[
\rho_s = \frac{\alpha_s \sum_{k=1}^K \pi_k ^s}{\beta_s (1-\sum_{k=1}^K \pi_k ^s)} .
\]
In our simulations, we impose $\rho_s >1$, with larger values corresponding to more modular hypergraphs. 
Note that in this setting, the total number of size-$s$ hyperedges is random and has expected value 
\[
\esp(|\calE_s|)= \binom{n}{s} \Big[\alpha_s \sum_{k=1}^K \pi_k ^s + \beta_s \Big(1-\sum_{k=1}^K \pi_k ^s\Big)\Big].
\]
We simulate hypergraphs with decreasing values $\esp(|\calE_s|)$ when $s$ increases, which is more realistic than the constant case.

\paragraph*{Hypergraphs with DCHSBM-like.}
We consider datasets simulated under the DCHSBM-like generating model proposed by \cite{Chodrow_21}. 
This model relies on a fixed true number of clusters $K$, balanced clusters  $|C_1^{\text{true}} |=\dots=|C_K^{\text{true}} |=n/K$ and equal  numbers of size-$s$ hyperedges for $2\le s \le S$; so that for each size $s$, a  total of $ |\calE|/(S-1)$ hyperedges are drawn. With probability $p_s$, such a hyperedge is placed on a $s$-tuple of (distinct) nodes within the same cluster and with probability $1-p_s$, it is placed on any $s$-tuple of (distinct) nodes.

The ratio $\rho_s$  of within-cluster over between-clusters size-$s$ hyperedges  is random and its expectation is \[
\esp(\rho_s)= \frac {p_s + (1-p_s) c_s}{(1-p_s)(1-c_s)} , \quad \text{ where }  c_s = \frac{K \binom{\lfloor n/K\rfloor }{s}} {\binom{n}{s}} = \frac{K (\lfloor n/K\rfloor ) ! (\lfloor n/K\rfloor -s)!}{n! (N-s)!}.
\]

Note that the DCHSBM-like generating model has been originally proposed for multisets hypergraphs, where nodes may be repeated in hyperedges, and hyperedges may be multiple. 
In practice, as we consider sparse hypergraphs where the number of hyperedges is linear wrt the number of nodes, multiple hyperedges are rare. 
Section~\ref{sec:SM_HSBM_DC} from the Appendix contains some further considerations on the links between parameters in HSBM and DCHSBM-like.

\paragraph*{h-ABCD benchmark dataset.}
Recently, \cite{h-abcd_paper} proposed a hypergraph artificial benchmark for community detection, called h-ABCD, together with a code for generating these modular hypergraphs \citep{h-abcd_code}. This generating model is an appropriate candidate to compare modularity approaches. In this model, we  fix the number of nodes $n$ and we either input a sequence of node degrees or it is sampled from a power-law distribution with some input exponent $\gamma\in (2,3)$ and input minimum/maximum degree values. The true clusters  sizes are also either input or sampled from a power-law with some input exponent $\beta \in (1,2)$ and minimum/maximum sizes values. 
In our case, we choose to fix the cluster sizes so that the number of clusters is fixed rather than random.
The model requires a sequence $q=(q_1,\dots, q_S)$ of weights summing to 1 such that $S$ is the maximal hyperedge size and $q_s$ is the fraction of size-$s$ hyperedges. For instance fixing $q_1=0$ prohibits self-loops. 
The script \texttt{abcdh.jl} also handles the proportion of homogeneous hyperedges, where homogeneity is the concept discussed in Section~\ref{sec:mod} when introducing $\Kawdc$. We recall that a homogeneous hyperedge has more than half of its nodes within the same (majority) cluster. 
Let $\omega_{c,s}$ denote the fraction of homogeneous hyperedges of size $s$ that have exactly $c\ge \lfloor s/2\rfloor $ nodes belonging to their majority cluster, so that  $\sum_{c=\lfloor s/2\rfloor +1} \omega_{c,s} = 1$. 
This notation is not to be confused with the weights $w_{s,c}$ introduced in \eqref{eq:wsc}. To link it to previously introduced  quantities, we remark that $\omega_{c,s} = \sum_{e \in \calE_s} \sum_{k=1}^K e_H^{s,c}(C^{\text{true}}_k)/|\calE_s|$. 
The current implementation handles 3 different options: \texttt{linear, strict, majority}, corresponding to the following choices 
\[
\omega_{c,s} = \left\{ 
\begin{tabular}{cc}
$(\lceil s/2 \rceil)^{-1}$ & if majority,\\
$\frac{2c 1\{c\ge s/2\}}{(s + \lfloor s/2\rfloor + 1)\lceil s/2\rceil}$ & if linear,\\
$1\{c=s\}$ & if strict.
\end{tabular}
\right.
\]
Thus in the \texttt{majority} setting, a homogeneous hyperedge is randomly drawn among all hyperedges with more than half of their nodes in their majority cluster, while in the \texttt{strict} setting, homogeneous hyperedges are exactly within-cluster hyperedges (i.e., all nodes belong to the majority cluster).  The \texttt{linear} setting spreads the homogeneous hyperedges in a linear fashion across the different values $c$ of the number of nodes in the majority cluster. In that setting, there is thus a larger number of homogeneous hyperedges showing a larger number of nodes in their majority cluster. 
Having  set which hyperedges are homogeneous ones, a mixing parameter $\xi\in (0,1)$ controls for the proportion  of the degree of each node that is assigned to non-homogeneous hyperedges. 
In this generating model, the total number of hyperedges is random and equals
\[
|\calE| = \frac{\sum_{v\in V} \deg_H(v)}{\sum_s s q_s}.
\]
In the strict setting, we can also express the ratio $\rho_s$ of within-cluster over between-cluster size-$s$ hyperedges as
\[
\rho_s = \frac{1-\xi}{\xi} \quad \text{for the strict setting of h-ABCD.}
\]

\begin{table}[htbp]
\centering
\begin{tabular}{C{1.6cm}C{8cm}p{4.5cm}}
\toprule
Synthetic model &Parameters & Characteristics \\
\hline
HSBM& $K$; clusters proportions $\pi$; within- and between-clusters probabilities $\alpha_{s}, \beta_{s}$  & random clusters sizes; random $|\mathcal E_2|,\dots,  |\mathcal E_S|$ \\
DCHSBM-like & $K$; $|\calE|$; probability $p_s$  that a hyperedge is placed on a $s$-tuple of within-cluster nodes  
& balanced clusters;  
equal nb of size-$s$ hyperedges $|\mathcal E_2|=\dots =|\mathcal E_S|= |\mathcal E| /(S-1) $\\
h-ABCD & degrees (power-law distribution or fixed values); cluster sizes (power-law distribution or fixed values); proportions $q_s$ of size-$s$ hyperedges; setting of  homogeneous hyperedges (majority, linear or strict); proportion $\xi$ of node degree assigned to homogeneous hyperedges
& random number and cluster sizes; random  $|\mathcal E_2|,\dots,  |\mathcal E_S|$\\
\bottomrule
\end{tabular}
\caption{Summary of synthetic models for modular hypergraphs and their characteristics. In all the models, the number of nodes $n$ and the maximal hyperedge size $S$ are chosen by the user. The numbers of clusters $K$ and hyperedges $|\calE| $ (resp. size-$s$ hyperedges $\calE_s$) can either be fixed or random. }
\label{tab:models}
\end{table}

\section{Experiments}
\label{sec:experiments}
\subsection{Scenarios}
\label{sec:scenario}

\paragraph*{General principles and base case scenario.}
Our simulations explore various settings in order to i) highlight the global behaviors of the methods and compare their performances; ii) explore which hypergraphs characteristics most impact those performances. In all the settings, we choose to focus on \emph{sparse} hypergraphs, for which the number of hyperedges grows linearly with the number of nodes, as this is a most realistic setting. 

We start with ``base case'' scenarios, called scenarios A, that we defined in the 3 different generating models (HSBM,  DCHSBM and h-ABCD). Then we explore other scenarios (called B to F and Z), simulated under the most convenient model to do so, and in which we modify only one characteristic at a time wrt the base case. 
Each scenario is composed of sub-cases with different samples sizes, comprising in general cases 1 to 6 corresponding to $n \in \{50, 100, 150, 200, 500,1000\}$. Moreover, in each scenario explored, we randomly generated 25 hypergraphs. 
Table~\ref{tab:characteristics} gives a summary of the scenarios considered and the empirical characteristics of the 25  hypergraphs generated for each of them.  In this table, each simulation is summarized  through its differing characteristic wrt the base case (namely, scenarios A). For example,  ScenB-DCHSBM is a simulation of hypergraphs less sparse than the base case.

\begin{center}
\begin{longtable}{C{4cm}|C{1.2cm}C{1.2cm}C{1.2cm}C{1.2cm}C{1.2cm}C{1.2cm}}

\hline Simulation &Scenario&  $n$ & $\bar{|\calE_2|}$ &$\bar{|\calE_3|}$ & 
$\bar d$ & max($d$) \\ \hline 
\endfirsthead

\multicolumn{7}{c}%
{{\bfseries \tablename\ \thetable{} -- continued from previous page}} \\
\hline Simulation &Scenario&  $n$ & $\bar{|\calE_2|}$ &$\bar{|\calE_3|}$ & 
$\bar d$ & max($d$) \\ \hline 
\endhead

\hline \multicolumn{7}{|r|}{{Continued on next page}} \\ \hline
\endfoot

\hline \hline
\endlastfoot

\textbf{ScenA-HSBM}& A1 & 50 & 198& 85 &13 &32\\
(base case)& A2 & 100 & 397& 178 &13 &26\\
  {$K=S=3$}& A3 & 150 & 592& 265 &13 &28\\
{balanced clusters}& A4 & 200 & 795& 354 &13 &28\\
$\rho_s=1.7$; $|\calE_2|/|\calE|\simeq 0.7$& A5 & 500 & 1990& 885 &13 &28\\
\hline
\textbf{ScenA-DCHSBM} & A1 & 50  & 194& 89 & 13&26\\
(base case)& A2 & 100  & 400& 174 &13 &29\\
 {$K=S=3$}& A3 & 150  & 604& 263 &13 &29\\
{balanced clusters}& A4 & 200  & 804& 345 &13 &32\\
{$\esp(\rho_s)=1.7$}& A5 & 500  & 2015&860  &13 &30\\
$|\calE_2|/|\calE|\simeq 0.7$& A6 & 1000  & 4030&1720  &13 &31\\
\hline
\textbf{ScenA-hABCD} & A1 & 50 &42 &12 &2.4  & 31 \\
(base case)& A2 & 100 &80 &24 &2.3  & 32 \\
 {$K=S=3$}& A3 & 150 & 117& 37&2.3  & 32 \\
{balanced clusters}& A4& 200 & 157& 51&2.3  & 32 \\
{$\rho_s=1.7$}& A5& 500 & 394& 132&2.3  & 32 \\
$|\calE_2|/|\calE|\simeq 0.75$& A6& 1000 & 782& 266&2.3  & 32 \\
\hline
\textbf{ScenB-DCHSBM} & B1 & 50  & 554&245  &37 &74\\
(less sparse)& B2 & 100  & 1117&483  &37 &61\\
{$K=S=3$}& B3 & 150  & 1680&720 &37 &59\\
balanced clusters& B4 & 200  & 2242&958  &37 &61\\
$\esp(\rho_s)=1.7$& B5 & 500  & 5614 &2386  &37 &63\\
$|\calE_2|/|\calE|\simeq 0.7$& B6 & 1000  &  11198&4802  & 37&62\\
\hline
\textbf{ScenC - HSBM} & C1 & 50 & 227& 100& 15 & 31\\
 (unbalanced clusters)& C2 & 100 &460 & 208&15  &37  \\
{$K=S=3$}& C3 & 150 & 690& 313& 15 &40  \\
$\pi=(1/6,1/3,1/2)$& C4& 200 &929 & 423&  15.5&35  \\
$\rho_s=1.7$& C5& 500 & 2319& 1063& 15.5& 39 \\
\hline
\textbf{ScenD-DCHSBM} & D1 & 50  & 84& 199  &15 &32\\
($|\calE_3| \gg |\calE_2|$)& D2 & 100  & 173&  402&15 &33\\
{$K=S=3$}& D3 & 150  & 258&607  & 16&31\\
balanced clusters& D4 & 200  &344&805  &16 &31\\
$\esp(\rho_s)=1.7$& D5 & 500  & 860& 2015 &16 &32\\
$|\calE_2|/|\calE|\simeq 0.3$& D6 & 1000  & 1722& 4028 &16 &36\\
\hline
\textbf{ScenE-DCHSBM} & E1 & 50  & 195& 88  & 13&28\\
(larger $\rho$)& E2 & 100  & 403& 172 &13 &26\\
{$K=S=3$}& E3 & 150  & 605& 262 & 13&27\\
balanced clusters& E4 & 200  &805&344  &13 &27\\
$\esp(\rho_s)=2$& E5 & 500  & 2008& 867  &13 &29\\
$|\calE_2|/|\calE|\simeq 0.7$& E6 & 1000  & 4040&1710  &13 &31\\
\hline
\textbf{ScenF-DCHSBM} & F1 & 50  &199 & 84  & 13&27\\
(smaller $\rho$)& F2 & 100  & 408&167  &13 &26\\
{$K=S=3$}& F3 & 150  &605 & 262 & 13&30\\
balanced clusters& F4 & 200  &811&334  &13 &27\\
$\esp(\rho_s)=1.4$& F5 & 500  & 2004& 871  &13 &31\\
$|\calE_2|/|\calE|\simeq 0.7$& F6 & 1000  & 4024&1726  &13 &30\\
\hline
\textbf{ScenZ-hABCD} & Z1 & 100 &49 &18 &1.5  & 10  \\
 (default h-ABCD) & Z2 & 150 &72 &27 &1.5  & 10 \\
$K$ random, $S=3$& Z3 & 200 & 96&37 &1.5  & 10  \\
unbalanced clusters& Z4& 500 & 239& 94&1.5  & 10 \\
linear setting& Z5& 1000 & 478& 187&1.5  & 10 \\
\hline

\caption{Simulation settings and empirical descriptors of the 25 simulated hypergraphs in each scenario (line): number of clusters ($K$), maximal hyperedge size ($S$),  within-cluster over between-clusters size-$s$ hyperedges ratio ($\rho_s$), number of nodes ($n$), mean number of size-$s$ hyperedges ($\bar{|\calE_s|}$), mean node degree ($\bar d$) and maximum node degree ($\max(d)$).}
\label{tab:characteristics} \\
\end{longtable}
\end{center}

We started from a first series of scenarios, called scenarios A, that play the role of a reasonable sparse case for the methods to work. To explore the robustness of our conclusions, these scenarios are presented under the 3 different generating models (HSBM,  DCHSBM and h-ABCD) relying on similar settings for sample size $n$ and number of hyperedges $|\calE|$.  
We set the numbers of hyperedges such that they grow linearly with the number of nodes $n$ (sparse setting). 
We generated $K=3$ clusters with equal size or probability (depending on the generating model) and  the maximum hyperedge size $S=3$.  
This latter choice ensures both reasonable computing times and  simplicity of model parametrization. 
The ratio $|\calE_2|/|\calE|=|\calE_2|/(|\calE_2|+|\calE_3|)$ is set to 0.7  (on average) to reflect the fact that we expect  larger sizes hyperedges to be less frequent than smaller-sizes ones. 
The within-cluster over between-cluster hyperedge ratio is constant wrt size $s\in \{2,3\}$ and set to $\rho_s=1.7$ (either exactly or on average), in order to obtain  modular hypergraphs.

For this scenario A, we first generated hypergraphs  under HSBM with a number of nodes $n$ up to 500, the algorithm becoming too slow for $n=1000$. 
Under DCHSBM, we went up to  sample size $n=1000$. 
Finally we generated samples under h-ABCD again up to a  number of nodes $n=1000$. In this latter model, we considered the strict setting regarding homogeneous hyperedges and choose the parameter $\xi$ such that the resulting $\rho_s=1.7$ and we set $|\calE_2|=3|\calE_3|$, which is approximately the case in the other 2 models.  
The degree distribution is scale-free with $\gamma=2.07$ and minimum  and maximum value set to 1 and 32, respectively (the observed values in the other 2 models). Note that the range of $\gamma\in (2,3)$ did not allow us to select mean degrees with similar values than with the other 2 models. 
In this sense, this scenario A under h-ABCD generating model diverges from the other ones (under HSBM and DCHSBM).

\paragraph*{Variant scenarios.} 
We further contrasted scenarios A by varying one characteristic at a time, keeping all others fixed. As our conclusions on scenarios A were globally robust against the choice of the generating model (at least among HSBM and DCHSBM, see next Section~\ref{sec:results}), we explored those variations in the most convenient model to do so. 
In scenario B, we  decrease the sparsity of the model by generating more hyperedges (keeping all other parameters identical as in scenario A).  
In scenario C, we explore the effect of unbalanced clusters, while in scenario D, we explore the effect of varying the proportions of size-2 and size-3 hyperedges, namely considering more size-3 than size-2 hyperedges.
Scenarios E (resp. F) considers the case where the  within-cluster over between-cluster hyperedge ratio $\rho_s$ is increased (resp. decreased) wrt scenario A. 
Finally, because we obtained pretty bad results for all modularity clustering methods relying on hypergraphs generated by h-ABCD (see next Section~\ref{sec:results}), we explored in scenario Z  the author's default values of that model to generate modular hypergraphs. Note that in this case, the true number of clusters $K$ is random and the ratio $\rho_s$ cannot be obtained from the model parameters.

\subsection{Quality assessment}
\label{sec:quality}
We now describe the different properties explored to assess the quality of each method. These properties are summarized in Table~\ref{tab:quality}. 

We first consider  accuracy of the clustering, relying on the Adjusted Rand Index \cite[ARI,][]{ARI} that measures similarity between $ \hat \calC$ and $\calC^{\textrm{true}}$ (up to label switching).  It is upper bounded by 1, where a value of 1 indicates perfect agreement between the clusterings, and negative values indicate less agreement than expected by chance. 
Then we consider running times (expressed in seconds) of each method. The results have been obtained on a computer with a AMD EPYC 7542 32-Core processor, 128 CPU (2 sockets of 32 double threads cores; we used just one core for each job as none of the procedure is parallelized) and 675Gb RAM. 
We already mentioned that modularity maximization is far from trivial because of the size of the search space. Thus, an important  question is whether the method at stake indeed maximizes its objective. To assess this, we measure the  relative error between the ground truth modularity $Q^{\textrm{true}}= Q(H, \calC^{\textrm{true}})$ and the resulting value $\hat Q= Q(H, \hat \calC)$ at the estimated classification $\hat \calC$, namely 
\[
\text{error}= \frac{Q(H, \calC^{\textrm{true}})-Q(H, \hat \calC)}{Q(H, \calC^{\textrm{true}})}.
\]
A method that reaches its objective (modularity maximization) without being able to recover the true modular clusters would reveal that it is based on a definition of modularity that is not appropriate. Also note that this error has a sign, with negative values indicating that ground truth modularity is not the maximum value. 
The mean values and standard deviations for ground truth modularity $Q^{\textrm{true}}$ are also reported (Table~\ref{tab:GT_mod} in Appendix), since values close to zero  could induce unstable errors. 

We finally also consider the estimated number of clusters $\hat K$ wrt its true value $K$. In general we present a barplot of the estimated values, to be compared to the true and fixed one. Only for scenarios Z where the true value $K$ is random, we plotted  the difference $\hat K-K$.

\begin{table}[htbp]
\centering
\begin{tabular}{C{7cm}C{8cm}}
\toprule
Question & Measure\\
\hline
Is the classification correct? & ARI($\hat \calC$ ; $\calC^{\textrm{true}})$ \\
Is the method fast ? & Running times \\
Is the modularity maximized? &  Relative error between  $Q^{\textrm{true}}$ and $\hat Q$ \\
Is the number of clusters correct ? & distribution of $\hat K $ wrt $K$\\  
\bottomrule
\end{tabular}
\caption{Quality assessment.}
\label{tab:quality}
\end{table}

\section{Results}
\label{sec:results}

\begin{figure}[htbp]
\centering
\includegraphics[width=\textwidth]{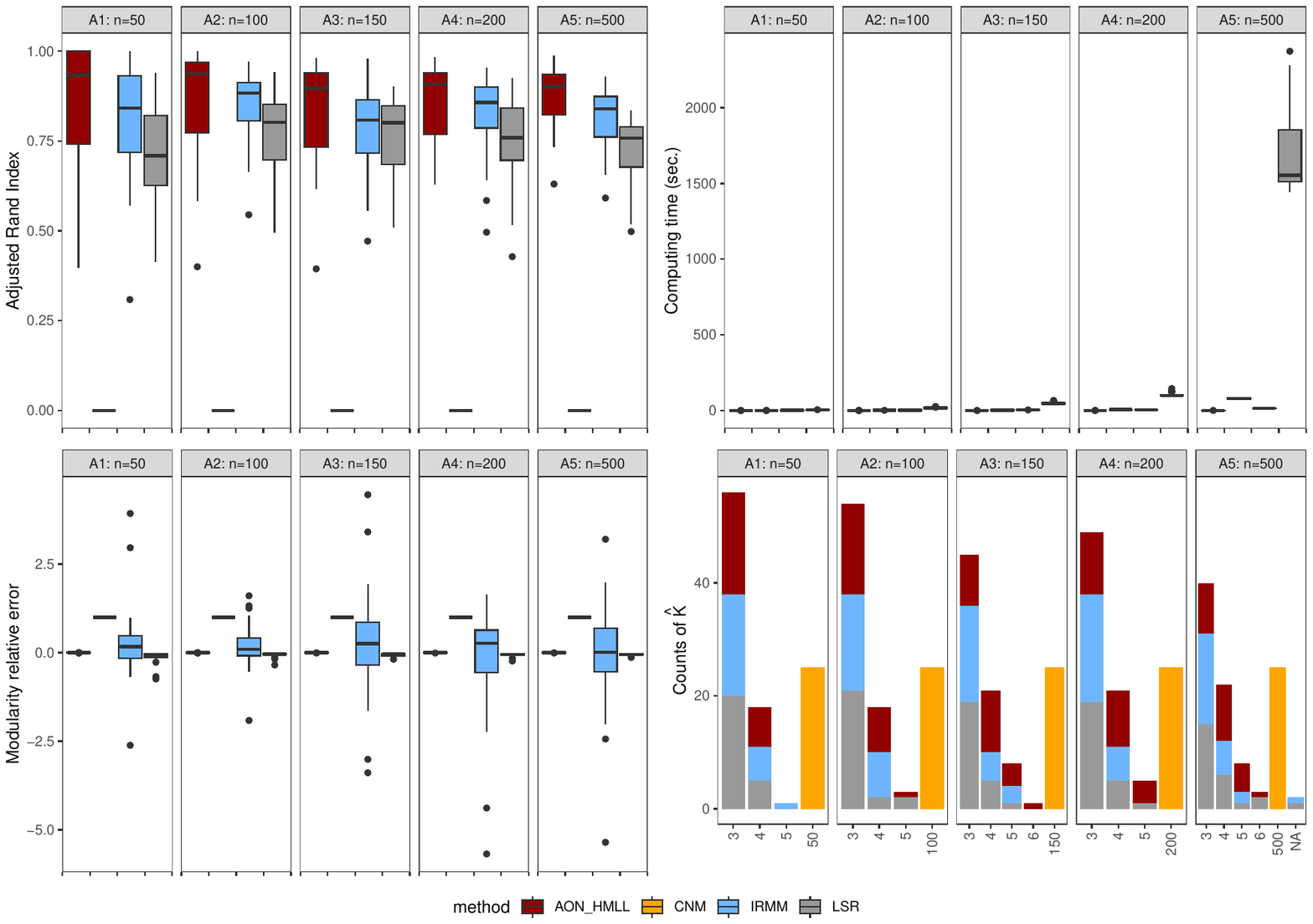}
\caption{Datasets HSBM, scenarios A1 to A5. Comparison by increasing the number of nodes from $n\in \{50,100,150,200,500\}$: Adjusted Rand Index (top left), time in seconds (top right),  relative error on  modularity (bottom left) and estimated number of clusters (bottom right, true value is $3$). The \texttt{IRMM} and (consequently) the \texttt{LSR} methods both gave an error on one dataset in scenario A5. 
Outlier points have been removed: from the  relative error plot (bottom left), 1 value below -500 concerning the \texttt{IRMM} method in scenario A1. Moreover, 
 one dataset from scenario A5 gave an error with the \texttt{IRMM} and (consequently) the \texttt{LSR} methods; corresponding results were removed from the plots.  
}
\label{fig:HSBM_A}
\end{figure}

\paragraph*{General comparison.}
We first analyze the results under the simplest scenarios (namely scenarios A, which represent our base case) and the HSBM generating model. 
Results are presented in Figure~\ref{fig:HSBM_A}. 
First, the \texttt{CNM-like} algorithm does not recover the ground truth clusters, with ARI values around 0 (Figure~\ref{fig:HSBM_A}, top left). In fact, the algorithm did not improve over its initialization at $\calC^{\text{own}}=(\{1\},\dots,\{n\})$ and the number of estimated clusters corresponds to the actual number of nodes (bottom right). Its  relative error on modularity is constant and corresponds to the  relative difference between the modularity of  $\calC^{\text{true}}$ and that of $\calC^{\text{own}}$. It is positive, so that the modularity maximization goal is clearly not achieved here.
The other 3 methods successfully recover the true clusters. 
For those 3 methods, median ARI values are above 0.7 (top left) and the number of estimated clusters  varies between 3 and 6 (bottom right). While the \texttt{AON-HMLL} globally obtains the best ARI results (top left), it is also the fastest method (top right) and it attains its objective of modularity maximization (relative error around 0, see bottom left). 
The \texttt{LSR} algorithm was proposed to improve over the \texttt{IRMM}. 
While  its  relative error on modularity (bottom left) seems in general improved over the latter (with  smaller values), 
Table~\ref{tab:GT_mod} in Appendix  shows that the modularity $\Ku$ optimized by  \texttt{IRMM} is close to 0 for the ground truth clusters, thus giving unstable errors; while $Q^{\textrm{linear}}$ optimized by \texttt{LSR} is strictly positive at those ground truth clusters. 
Most importantly,  from the clustering point of view, ARI is not improved (top left) and computing times are much larger (top right). 
This seems to indicate that the \texttt{LSR}  places too much emphasis on maximizing modularity at the expense of clustering recovery.
As the number of nodes $n$ increases, we observe that ARI values globally have a lower dispersion, but  do not seem to overall improve (top left). This might be due to our setting where the within-cluster over between-cluster hyperedge ratio $\rho_s$ is kept constant when $n$ varies.

\begin{figure}[htbp]
\centering
\includegraphics[width=\textwidth]{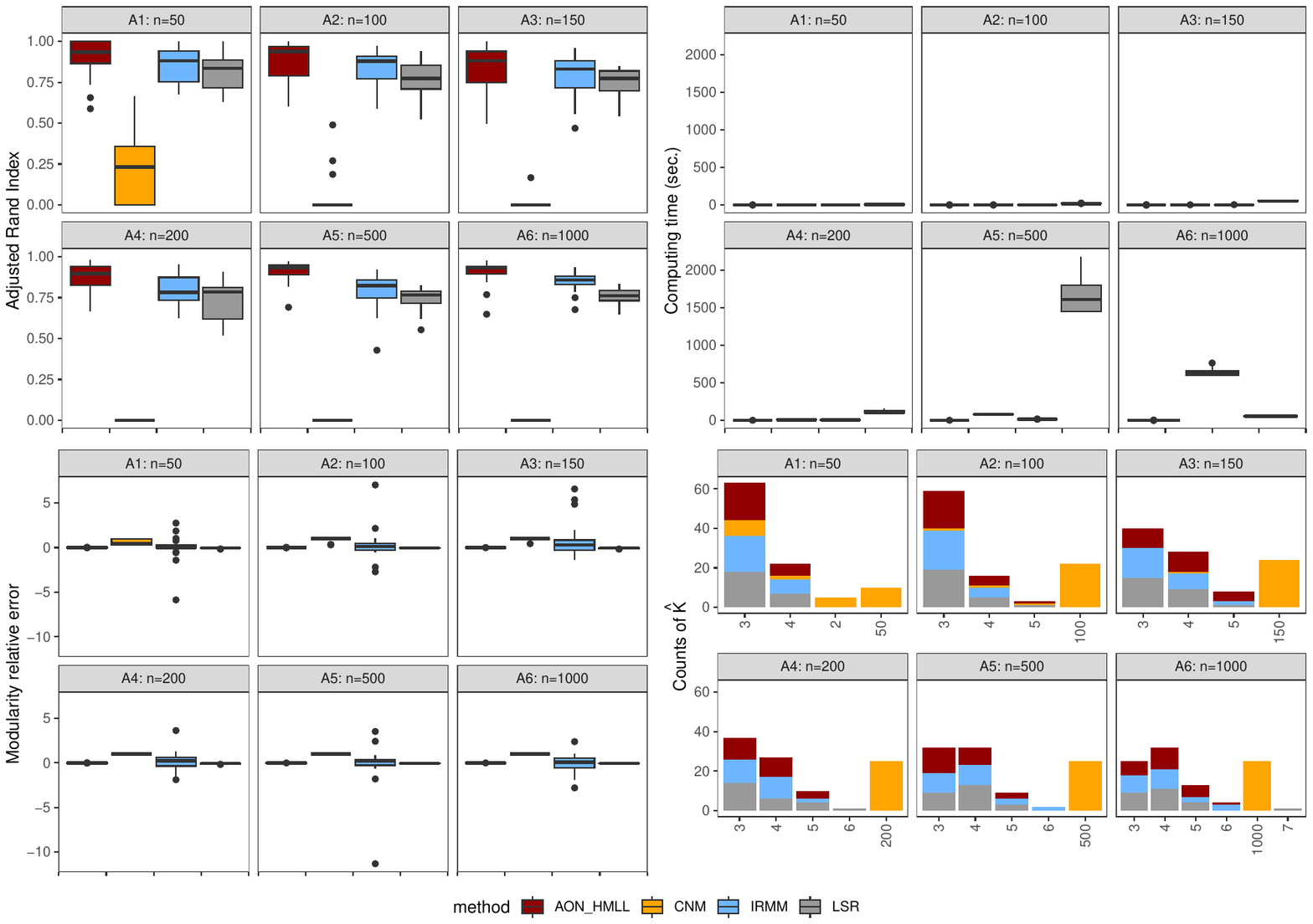}
\caption{Datasets DCHSBM, scenarios A1 to A6. Comparison by increasing the number of nodes from $n\in \{50,100,150,200,500,1000\}$: Adjusted Rand Index (top left), time in seconds (top right),  relative error on  modularity (bottom left) and estimated number of clusters (bottom right, true value is $3$). 
From the time plot (top right),  values for the \texttt{LSR}  method in scenario  A6 range between 15,796 and 22,350 seconds and are not shown. 
Outlier points have been removed from the  relative error plot (bottom left): 1 value above 300 concerning the \texttt{IRMM} method in scenario A4. }
\label{fig:DCHSBM_A}
\end{figure}

Let us now compare these results with those obtained on scenarios A generated under DCHSBM and presented in Figure~\ref{fig:DCHSBM_A}. 
From these simulations, we  confirm the previous conclusions: the \texttt{AON-HMLL} is globally the best method and  the \texttt{CNM-like} algorithm has very low performance for clustering recovery (ARI values very small). The other 2 methods successfully recover the clusters but the \texttt{LSR} does not improve on the \texttt{IRMM} and has a much larger computing time. Computing times are similar in this simulation and the former one; to see this, we choose to remove computing times for the \texttt{LSR} method in scenario A6 (Figure~\ref{fig:DCHSBM_A}, top right). Indeed, those values are all above 15,000 seconds and including them would have changed the $y$-scale in a way preventing from any possible comparison.
As a consequence, we conclude that our analysis is robust against the choice of HSBM or DCHSBM generating model.

\begin{figure}[htbp]
\centering
\includegraphics[width=\textwidth]{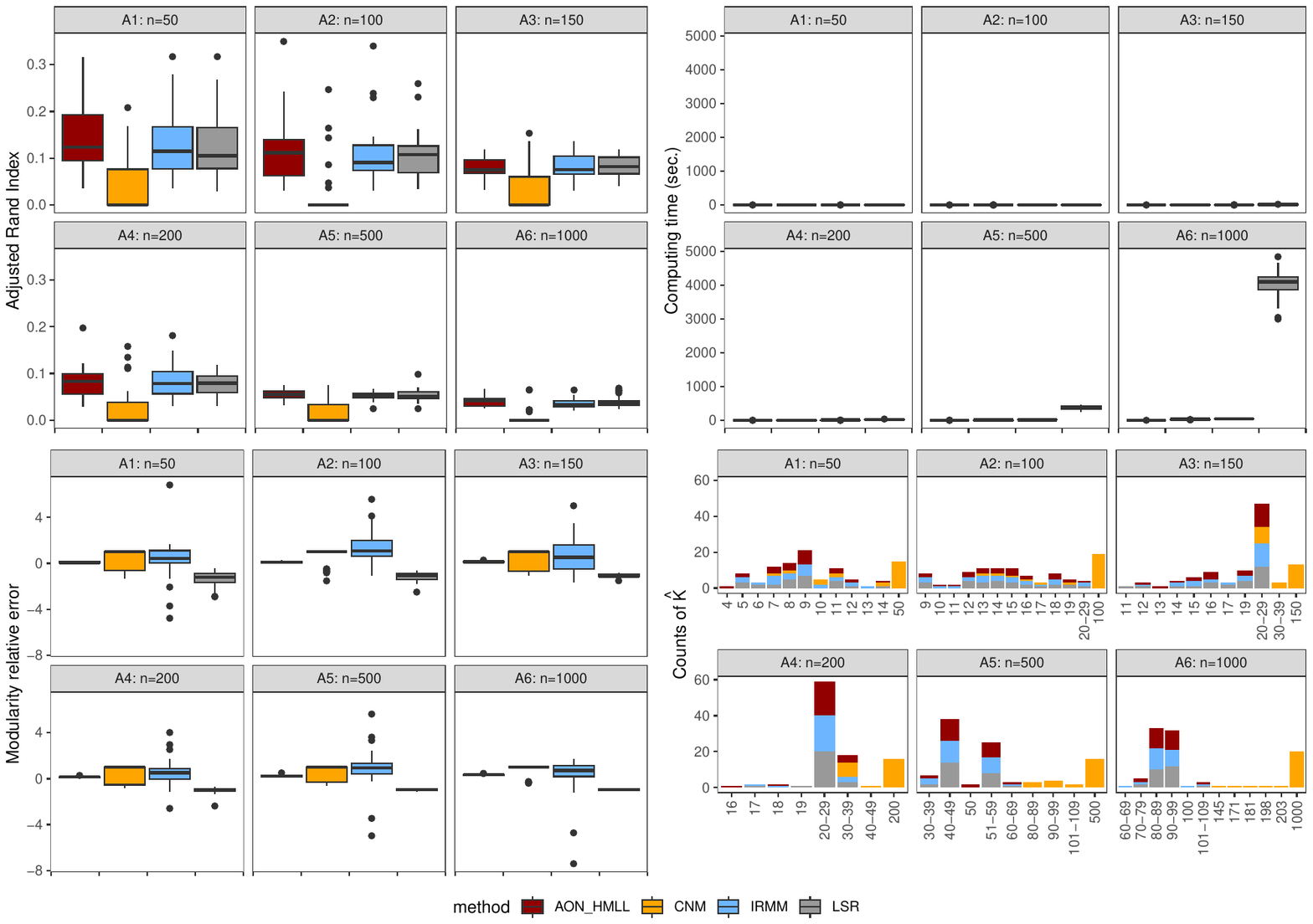}
\caption{Datasets h-ABCD, scenarios A1 to A6. Comparison by increasing the number of nodes from $n\in \{50,100,150,200,500,1000\}$: Adjusted Rand Index (top left), time in seconds (top right),  relative error on  modularity (bottom left) and estimated number of clusters (bottom right, true value is $3$). Outlier points have been removed: from the  relative error plot (bottom left), 3 values at 25, -50 and -55  concerning the \texttt{IRMM} method with in scenarios A6, A2 and A3 respectively. }
\label{fig:hABCD_A}
\end{figure}

To finish with these settings from scenarios A, we consider Figure~\ref{fig:hABCD_A} where the results for hypergraphs generated under the h-ABCD benchmark method are provided. Let us recall that while we tried to mimic as much as possible the characteristics of the scenarios A obtained under HSBM and DCHSBM, it was impossible to obtain similar node degrees within that h-ABCD generating process (see Table~\ref{tab:characteristics}) and the ones obtained here are much smaller. We observe that in this setting, none of the proposed methods is able to reconstruct the true clusters: ARI values are generally lower than 0.3 (see Figure~\ref{fig:hABCD_A}, top left) and the number of estimated clusters is too large (bottom right). Nonetheless, the modularity maximization seems to work as the  relative error between the ground truth modularity and its estimation is small (bottom left). Note also that the \texttt{LSR} algorithm seems to find a clustering with larger value of the $Q^{\textrm{linear}}$ modularity than at the ground truth clusters (negative errors).  
Overall, our conclusions raise the following question: are these datasets indeed modular?  We will come back to this later when discussing scenarios Z. 

We now explore additional insights on the methods performances provided by other scenarios. 

\begin{figure}[htbp]
\centering
\includegraphics[width=\textwidth]{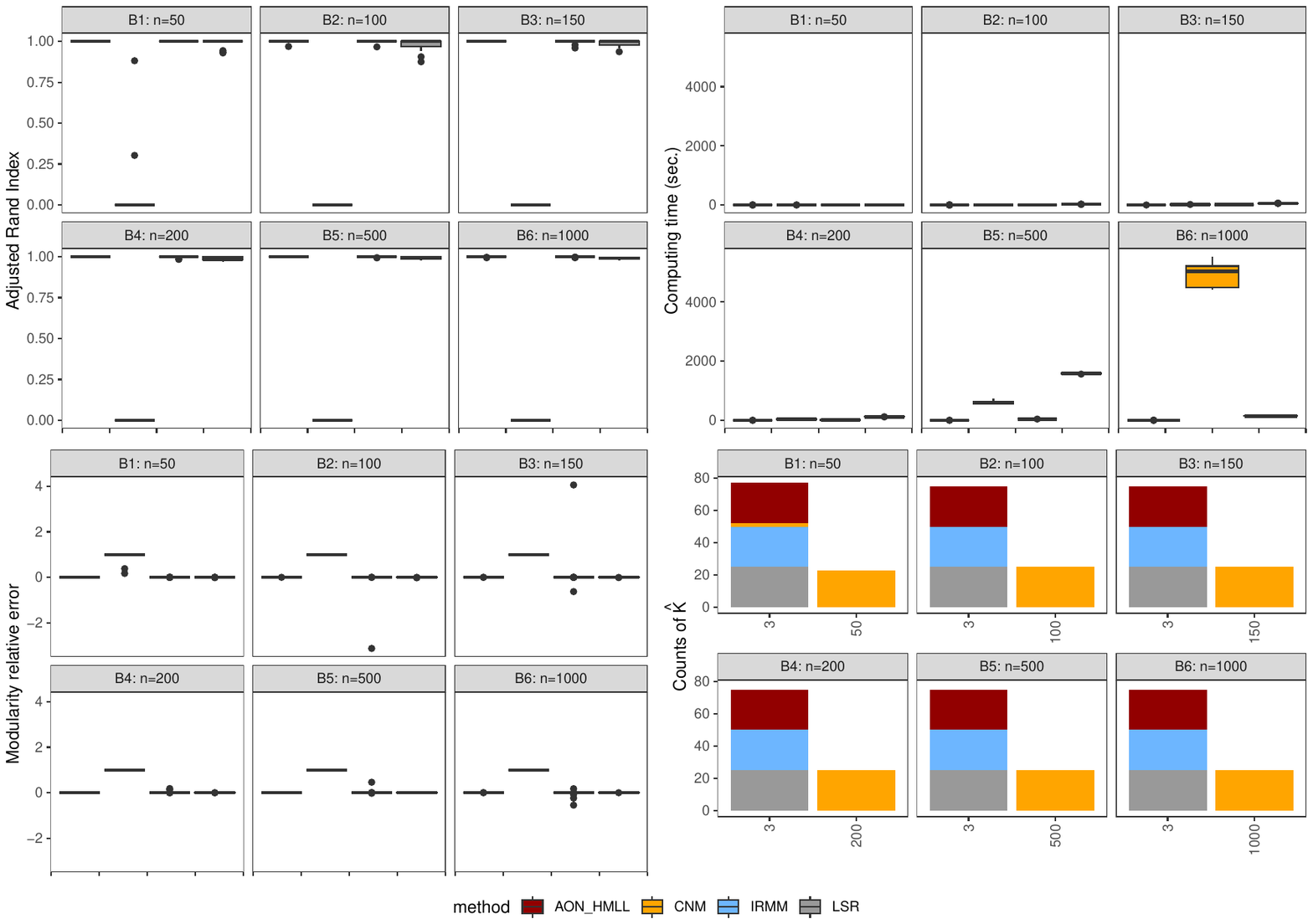}
\caption{Datasets DCHSBM, scenarios B1 to B6.  Comparison by increasing the number of nodes from $n\in \{50,100,150,200,500,1000\}$: Adjusted Rand Index (top left), time in seconds (top right),  relative error on  modularity (bottom left) and estimated number of clusters (bottom right, true value is $3$). 
From the time plot (top right),  values for the \texttt{LSR}  method in scenario  B6 range  between 16,961  and  17,895 seconds  are not shown. 
}
\label{fig:DCHSBM_B}
\end{figure}

\paragraph*{Impact of sparsity.}
In scenario B, we decreased the sparsity wrt to scenario A (note that the hypergraphs remain nonetheless sparse, see Table~\ref{tab:characteristics}). Results are presented in Figure~\ref{fig:DCHSBM_B}. Here again, we removed from the time plot (top right) all values for the \texttt{LSR} method in scenario B6. Their range between   16,961 seconds and  17,895 seconds would have changed the $y$-scale. 
 We mostly observe that while the above conclusions are still valid, the performances of  the 3 ``working'' methods (\texttt{AON-HMLL, IRMM} and \texttt{LSR}) increase wrt to scenario A. Indeed, except for the \texttt{CNM-like} algorithm, the methods exactly recover the true number of clusters (bottom right) and ARI values are almost equal to 1 (top left).  Relative errors on modularity are also almost zero for those 3 methods, indicating that the local maximization of the modularity works. 
We note that the \texttt{CNM-like} method has  relative error equal to 1. This comes from the fact that the maximized modularity is zero while the ground truth modularity is not zero. Also note that the computing time for this method in scenario B6 becomes significantly larger.

\begin{figure}[htbp]
\centering
\includegraphics[width=\textwidth]{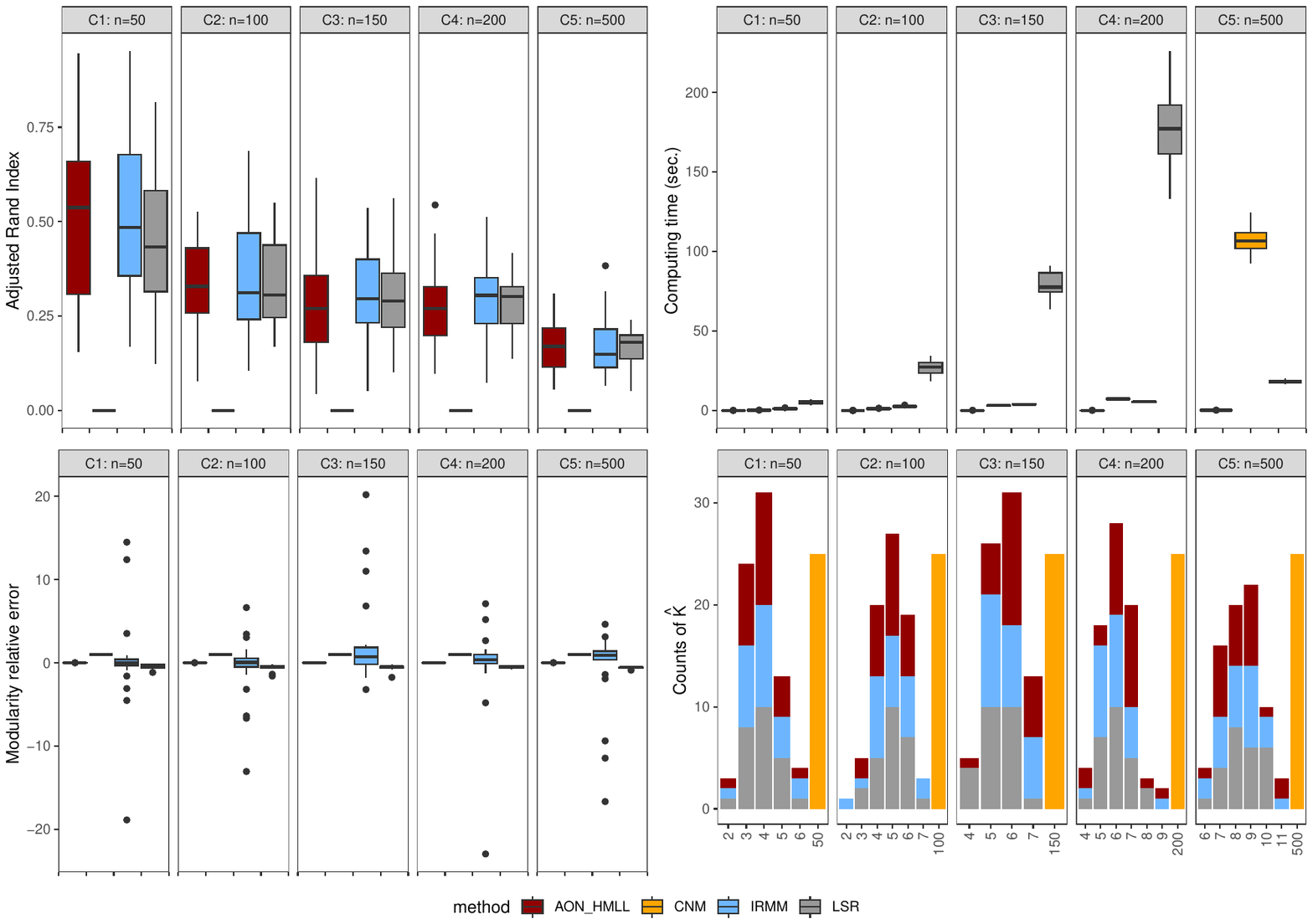}
\caption{Datasets HSBM, scenarios C1 to C5. Comparison by increasing the number of nodes from $n\in \{50,100,150,200,500\}$: Adjusted Rand Index (top left), time in seconds (top right),  relative error on  modularity (bottom left) and estimated number of clusters (bottom right, true value is $3$). 
From the time plot (top right),  values for the \texttt{LSR}  method in scenario  C5 range between  2,646 and  3,842 seconds and are not shown. 
Outlier points have been removed: 
from the  relative error plot (bottom left), 2 values concerning the \texttt{IRMM} method, one above 30 in scenario C3 and the second below -60 in scenario C4. }
\label{fig:HSBM_C}
\end{figure}

\paragraph*{Impact of unbalanced clusters.}
Let us now turn to scenario C where we explore the impact of unbalanced clusters. Results are presented in Figure~\ref{fig:HSBM_C}, where  we removed from the time plot (top right) all values for the \texttt{LSR} method in scenario C5 as they range between 
2,646 and  3,842 seconds. 
We observe that the overall performances of the methods have decreased wrt scenario A: ARI values are quite low (top left) and the number of clusters is over-estimated (bottom right). Contrarily to scenario A, increasing the number of nodes $n$ degrades the performance of ARI. This is quite counter intuitive, as we expect that with larger values of $n$, the clusters sizes increase and thus should be easier to detect.

\begin{figure}[htbp]
\centering
\includegraphics[width=\textwidth]{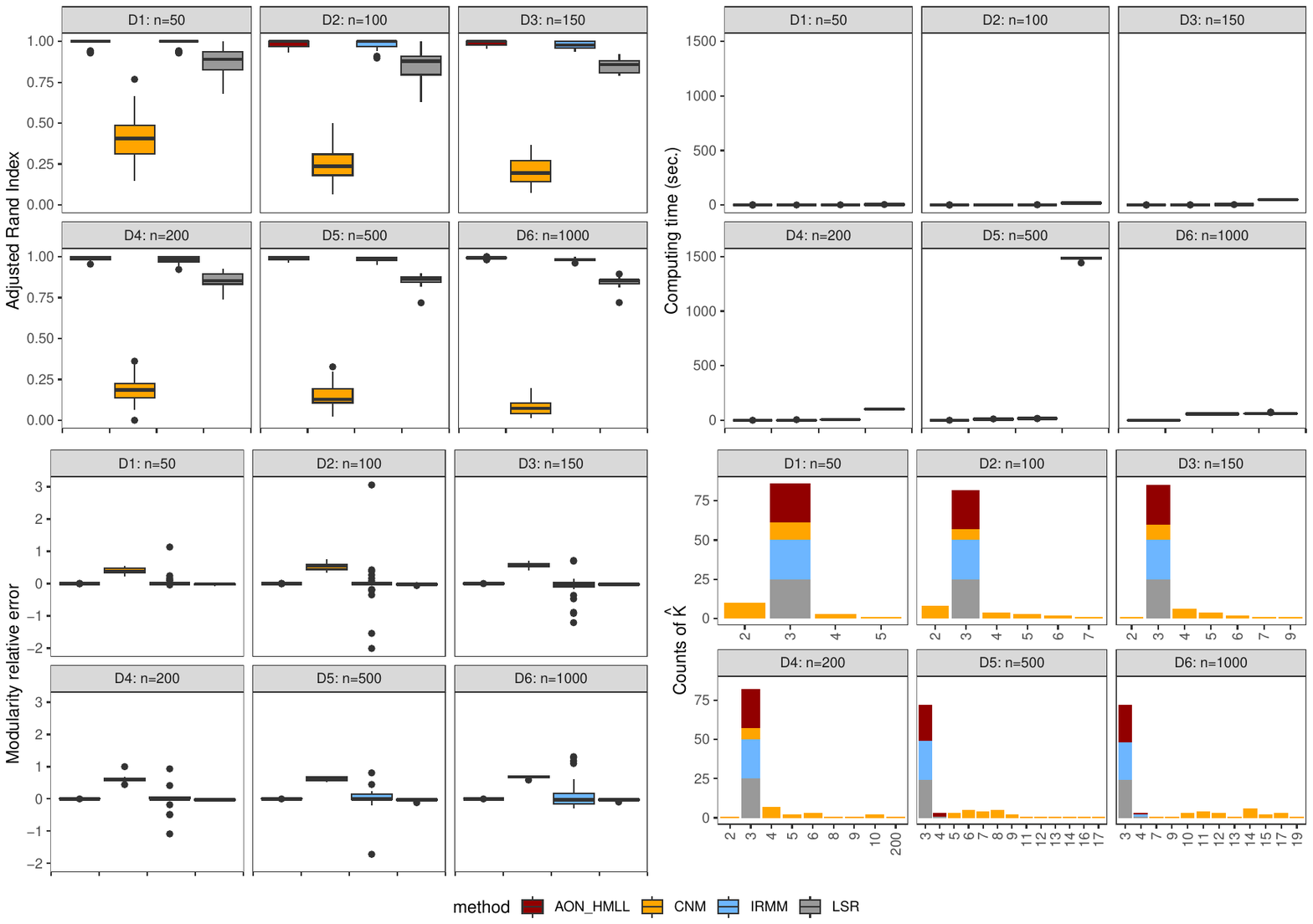}
\caption{Datasets DCHSBM, scenarios D1 to D6. Comparison by increasing the number of nodes from $n\in \{50,100,150,200,500,1000\}$: Adjusted Rand Index (top left), time in seconds (top right),  relative error on  modularity (bottom left) and estimated number of clusters (bottom right, true value is $3$). 
From the time plot (top right),  values for the \texttt{LSR}  method in scenario  D6 range between  17,153 and  22,377 seconds and are not shown. 
Outlier points have been removed: 
from the time plot (top right), 1 value above 2,800 seconds concerning the \texttt{LSR} method in scenario D5 and from the  relative error plot (bottom left): 1 value larger than 14 and 1 smaller than -11 concerning the \texttt{IRMM} method in scenario D5. }
\label{fig:DCHSBM_D}
\end{figure}

\paragraph*{Impact of proportions of  size-$s$ hyperedges.}
In scenario D, we explore the impact of the proportions of size-$s$ hyperedges. More precisely, while scenario A relied on a realistic setting of a smaller  number of size-$3$ than size-2 hyperedges (namely, $|\calE_2| \gg |\calE_3|$), we explore here the converse setting where $|\calE_3| \gg |\calE_2|$.  Results are presented in Figure~\ref{fig:DCHSBM_D}, where again, we removed from the time plot (top right), all  values for the \texttt{LSR}  method in scenario  D6 for comparison purposes, as their range is between  17,153 and 22,377 seconds. 
Here, we observe as in scenario B that the performances of  the 3  methods \texttt{AON-HMLL, IRMM} and \texttt{LSR} increase wrt to scenario A. Indeed,  the \texttt{AON-HMLL} and the \texttt{IRMM} have ARI values equal or close to 1 (top left) and find (almost always) the correct number of clusters (bottom right), indicating (almost) perfect clustering recovery. 
 Relative errors on modularity are also almost zero for those 3 methods, indicating that the local maximization of the modularity works. 
From that simulation we conclude that clustering via modularity maximization is easier for datasets with a larger proportion of large-size hyperedges and conversely, more difficult in the realistic setting where larger sizes hyperedges are in smaller proportion.

\begin{figure}[htbp]
\centering
\includegraphics[width=\textwidth]{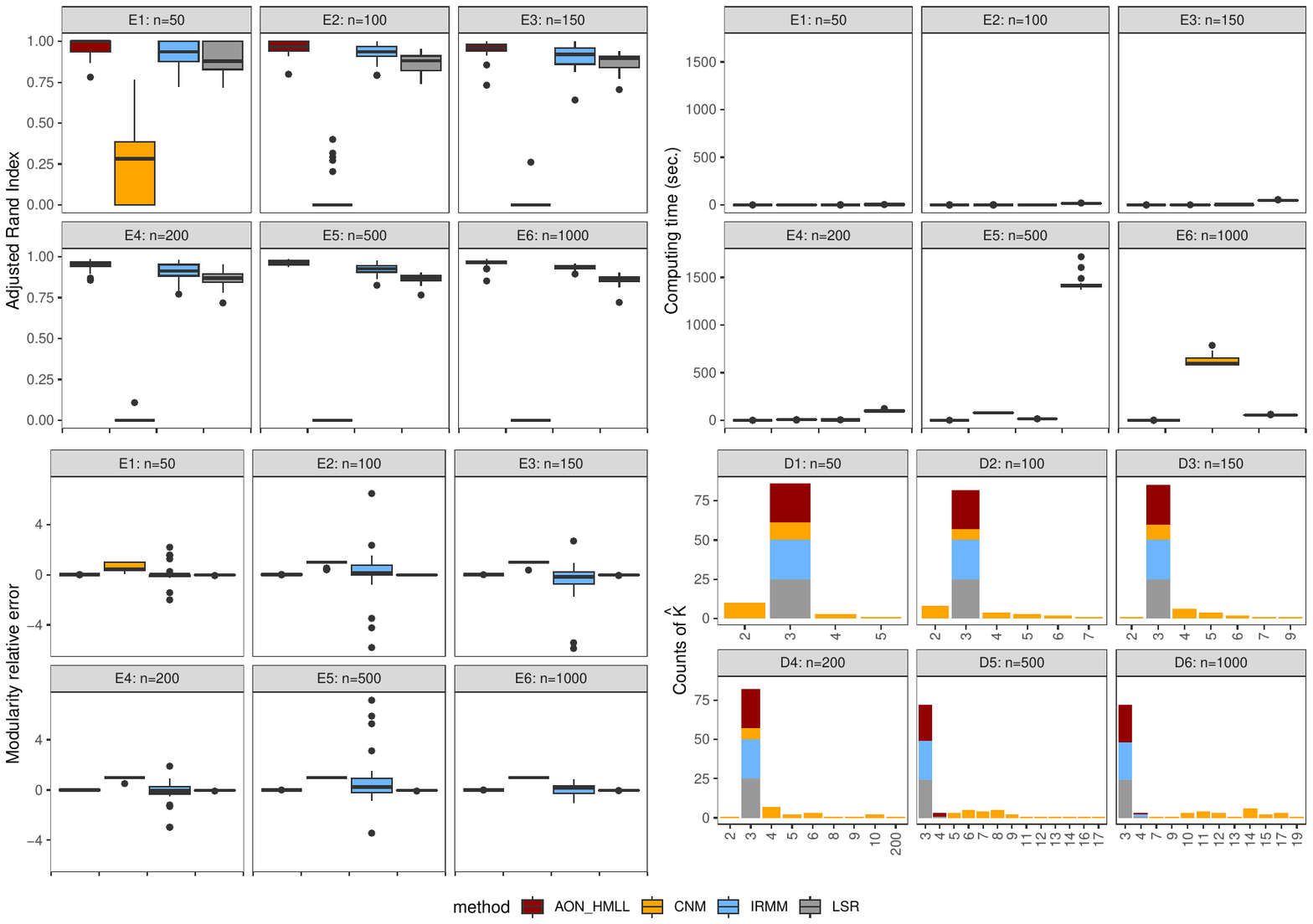}
\caption{Datasets DCHSBM, scenarios E1 to E6. Comparison by increasing the number of nodes from $n\in \{50,100,150,200,500,1000\}$: Adjusted Rand Index (top left), time in seconds (top right),  relative error on  modularity (bottom left) and estimated number of clusters (bottom right, true value is $3$). 
From the time plot (top right), values for LSR method in scenario E6 range between 15,833 and 20604 seconds and are not shown.
Outlier points have been removed from the  relative error plot (bottom left): 2 values concerning the \texttt{IRMM} method, one larger than 100 in scenario E3 and the other smaller than -21 in scenario E4. 
}
\label{fig:DCHSBM_E}
\end{figure}

\begin{figure}[htbp]
\centering
\includegraphics[width=\textwidth]{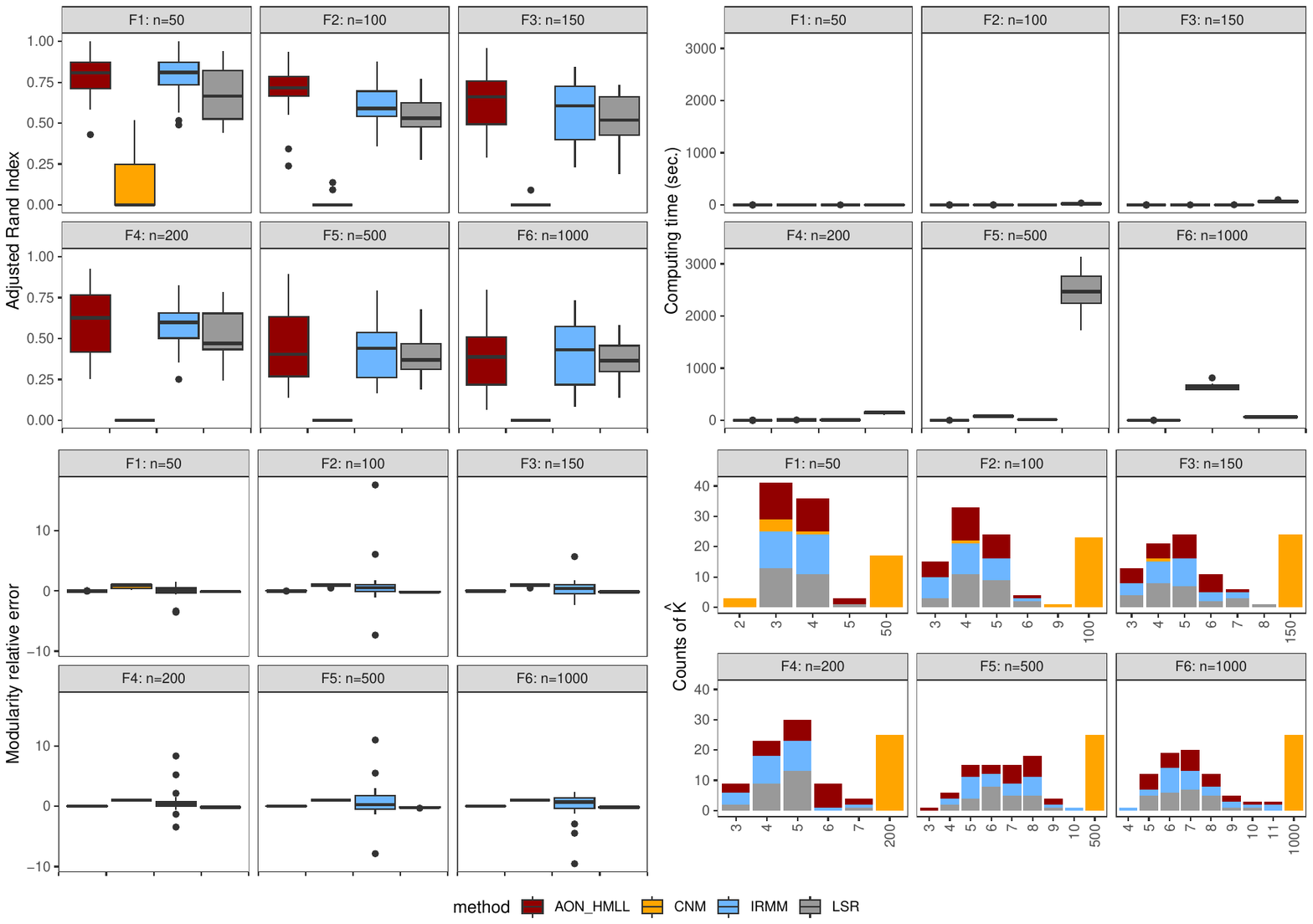}
\caption{Datasets DCHSBM, scenarios F1 to F6. Comparison by increasing the number of nodes from $n\in \{50,100,150,200,500,1000\}$: Adjusted Rand Index (top left), time in seconds (top right),  relative error on  modularity (bottom left) and estimated number of clusters (bottom right, true value is $3$). 
From the time plot (top right), values for LSR method in scenario F6 range between 22,891 and  39,967 seconds and are not shown.
Outlier points have been removed from the  relative error plot (bottom left): 3 values concerning the \texttt{IRMM} method, with 2 values smaller than -40 and -680 and 1 value larger than 22 in scenarios F1, F4 and F6 respectively.  
}
\label{fig:DCHSBM_F}
\end{figure}

\paragraph*{Impact of within-cluster over between-cluster hyperedges ratio.}
Scenario E (resp. F) rely on a larger (resp. smaller) value for the  within-cluster over between-cluster hyperedge ratio $\rho_s$ (still constant with hyperedge size $s$) compared to scenario A. The results of this simulation are presented in Figure~\ref{fig:DCHSBM_E} (resp. Figure~\ref{fig:DCHSBM_F}). 
In those figures again, time values for the \texttt{LSR} method in scenarios E6 and F6 respectively have been removed. 
We can observe that the modularity based methods are sensitive to this parameter $\rho_s$, with better clustering results obtained when this ratio is large. As expected, the more modular the hypergraphs are, the easier it is to recover the clusters.

\begin{figure}[htbp]
\centering
\includegraphics[width=\textwidth]{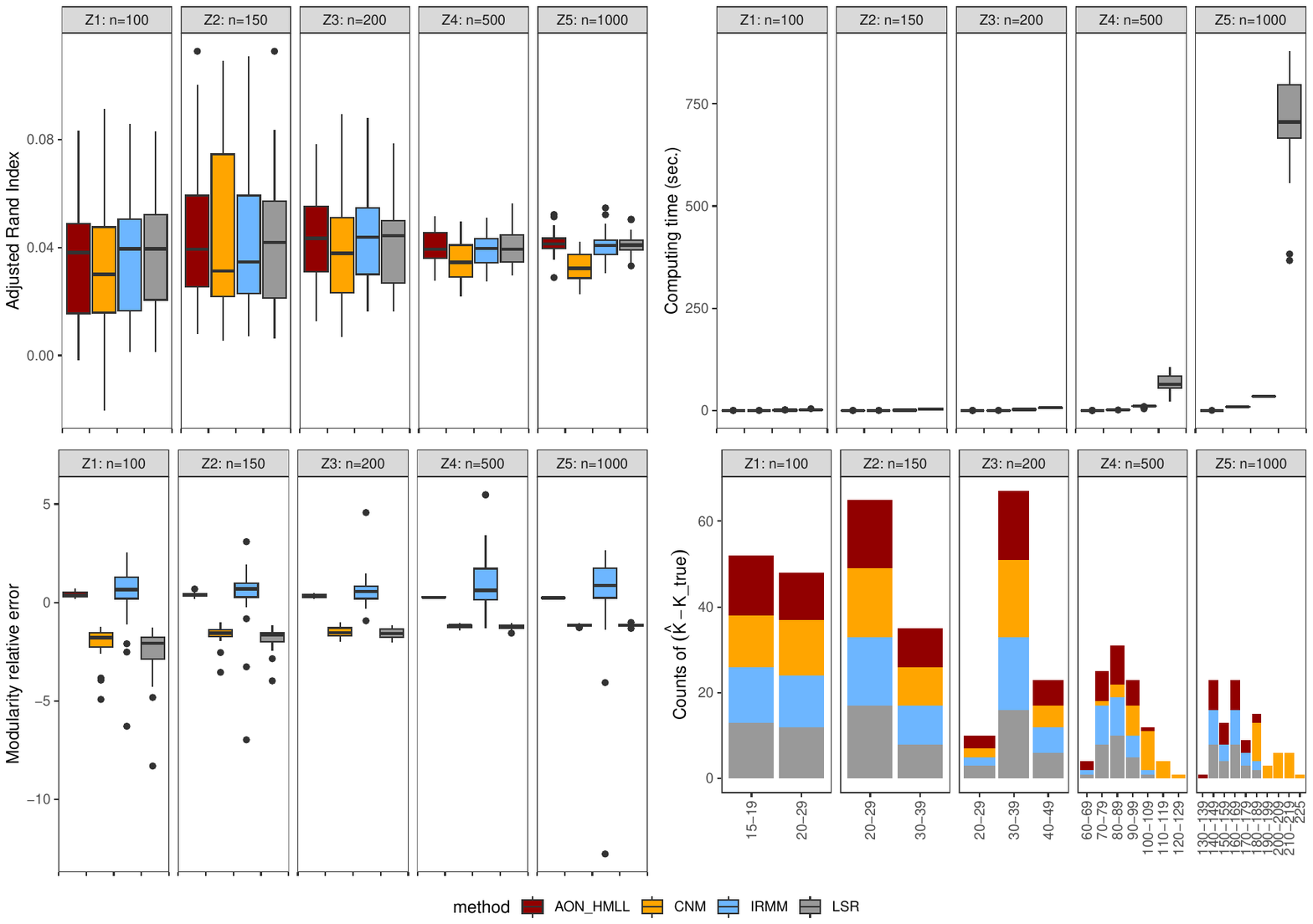}
\caption{Datasets h-ABCD, scenarios Z1 to Z5. Comparison by increasing the number of nodes from $n\in \{100,150,200,500,1000\}$: Adjusted Rand Index (top left), time in seconds (top right),  relative error on  modularity (bottom left) and difference between estimated number of clusters and true value (bottom right). Outlier points have been removed: from the time plot (top right), 4 values above 900 seconds concerning the \texttt{LSR} method  in scenario Z5  and  from the  relative error plot (bottom left), 1 value below -28 concerning the \texttt{IRMM} method  in scenario Z4. }
\label{fig:hABCD_Z}
\end{figure}

\paragraph*{Exploring possible bias from generating models.}
The bad results obtained by all methods on the datasets generated from scenarios A under h-ABCD model raised the question whether those hypergraphs are indeed modular. As we choose the settings of this simulation to mimic the observations obtained under HSBM and DCHSBM but did not completely succeed in that task, one could wonder whether our parameter choices make sense for this model. That is why we consider  scenarios Z under h-ABCD, relying on the authors of the model default parameter choices. Note that we started at sample size $n=100$ because $n=50$ did not work. 
The results obtained on these datasets are presented in Figure~\ref{fig:hABCD_Z}. Here, we observe that again, none of the methods is able to recover the ground truth clusters (top left plot shows ARI values around 0 and bottom right plot shows difference between estimated and true number of clusters quite large). This seems to indicate that h-ABCD is not an appropriate  benchmark method to test community detection algorithms.

Overall,  we could wonder whether the generating models DCHSBM and HSBM could be  favoring the \texttt{AON-HMLL} method. This could be particularly the case for the DCHSBM model as this model and the \texttt{AON-HMLL} method both derived from the same article \citep{Chodrow_21}. However, we can argue against that claim that the modularities $\ChAON$ and $\Ka$ maximized  by the methods \texttt{AON-HMLL} and \texttt{CNM-like}, respectively (see summary in Table~\ref{tab:algos}) both focus on contributions by within-clusters hyperedges only. More precisely,  the difference in  $\ChAON$ and $\Ka$ lies only on adaptive weights included in the former, the latter appearing as a special choice of those weights. We thus conclude that our simulations that partly focused on the ratio of within-clusters over between-clusters hyperedges is not especially in favor of the \texttt{AON-HMLL} method.\\

As a final note, we notice that in our experiments, the \texttt{IRMM} method sometimes shows some very large values for the  relative modularity error (points that we called ``outliers'' and removed to preserve $y$-scales in the plots). Looking at Table~\ref{tab:GT_mod} in the Appendix we observe that the ground truth modularity $\Ku$ is close to zero, explaining this unstable behaviour of the relative error.

\section{Discussion}
\label{sec:discu}
Let us now summarize the main findings of this study:
\begin{itemize}
\item  Globally, the best modularity-based approach is the \texttt{AON-HMLL}, as it often recovers the ground truth clusters and is among the fastest approaches; 
\item The \texttt{IRMM} algorithm has often good results at recovering ground truth clusters, but it is less fast than the \texttt{AON-HMLL};
\item Though the  \texttt{LSR} algorithm is specifically designed to improve on the \texttt{IRMM}, it does not improve the clustering problem at stake;
\item The \texttt{CNM-like} algorithm does not  recover the ground truth clusters in any simulation setting; 
\item We did not observe any algorithm for which the modularity $Q$ would be correctly maximized (relative  error in modularity close to zero) while clusters would not be recovered (low ARI values). Nonetheless, the modularity $\Ka$ from \cite{kami:etal:19} is not fully maximized by the \texttt{CNM-like} method, which leaves open the question of whether it is able to capture communities in hypergraphs. 
\end{itemize}

In the following, we concentrate on commenting the results of the ``working methods'', namely the  \texttt{AON-HMLL, IRMM} and the \texttt{LSR}: 
\begin{itemize}
\item The working methods tend to have better results when the densities of the hypergraphs increase, though still in a  sparse setting (i.e., when the number of hyperedges increases, see scenarios B);
\item The methods are sensitive to the balance in the cluster sizes, with better results when clusters are balanced (see scenarios C); 
\item  The methods tend to have better results when we observe a larger proportion of larger-size hyperedges (i.e., when $|\calE_3|$ becomes larger than $|\calE_2|$, see scenarios D);
\item The methods are sensitive to the ratio $\rho_s$ of within-cluster over between-cluster size-$s$ hyperedges, with better results when this ratio is larger (thus the hypergraph is more modular, see scenarios E and F).
\end{itemize}
Another conclusion from our study is that the h-ABCD benchmark model \citep{h-abcd_paper} does not seem appropriate to generate modular hypergraphs, or at least that none of the current modularity-based approaches is able to detect the simulated clusters in those hypergraphs. \\

Our work is a first building block  in gaining a better understanding of modularity in hypergraphs, yet it comes with certain limitations that warrant attention in future research. One constraint arises from computational limitations  in both the generating models and modularity maximization methods, restricting our exploration to relatively small graphs (with a number of nodes $n\leq 1,000$). Consequently, we constrained ourselves to a limited number of clusters ($K\le 3$), as larger values might lead to clusters too small for effective detection. 
Our focus was on binary hypergraphs, which already encompass a vast array of higher-order interactions. However, weighted hypergraphs are also of significant interest. 
Additionally, our approach relied on simulated hypergraphs with characteristics dictated by methodological constraints (e.g., the number of nodes, number of clusters) and others chosen to align with what we believe to be realistic (e.g., sparse hypergraphs, $|\calE_2| \gg |\calE_3|$, $\dots$).  \cite{hg_char} examined 13 real-world hypergraphs with heterogeneous sparsity (ratios $|\calE|/n$ ranging from as small as 0.5 to around 50) and an average hyperedge size $s$  generally less than 3.9, with two exceptions (hypergraphs related to drug chemicals). Despite this attempt, the literature still lacks a large-scale study on the characteristics of real-world hypergraphs that could inform and support simulations.

There remain numerous unresolved questions that extend beyond the scope of the present contribution. Realistic characteristics, influenced by parameter choices in the generating models, are intricately tied to the issue of detectability thresholds. Specifically, under what circumstances is it possible to effectively recover clusters in a hypergraph? While this question has garnered attention for uniform hypergraphs \citep{ange:etal:15,Chien_etal19,step:zhu:22,Zhang_Tan_23}, real-world hypergraphs, which are non-uniform, remain largely unexplored in this context.
Furthermore, moving beyond clustering recovery, it would be valuable to investigate the discriminative power of modularities. Specifically, understanding how discriminative each proposed modularity measure is could provide insights on their design. Examining the distribution of modularity values across a diverse set of hypergraphs, including non-modular ones, holds significant importance.  
In a similar vein,  whether hypergraph  modularities are unimodal or not is an important question. 
Characterizing the behavior of modularities across the entire spectrum of node clusterings would aid in designing suitable modularity-based methods for community detection in hypergraphs.

\section{Availability}
\label{sec:available}
All the material for reproducing the simulations can be found online at \url{https://github.com/veronicapoda/modularity/}.

\section*{Acknowledgements}
The authors would like to thank Professor Veronica Vinciotti for fruitful discussions and her comments on earlier versions of this work.


\newpage
\appendix

\section{A symmetric hypergraph modularity}
\label{sec:SM_Chsym}
\cite{Chodrow_21} define  a \emph{symmetric} modularity, where for any partition $\calC$ of the set of nodes, the contribution of a  hyperedge $e\in \calE$ to the modularity of this partition is characterized  only by the vector $\bp$ whose   entries $p_k$ count the number of nodes in $e$ belonging to the $k$-th largest part in $e\cap \calC$.
More precisely, fix $\calC=(C_1,\dots,C_K)$  a partition of $V$ into $K$ parts and consider $e\subset V$ a subset of nodes (here, $e$ is not necessarily a hyperedge). Nodes in $e$ are partitioned by $\calC$ into parts, namely $e\cap \calC=(e_1,\dots,e_{J_e})$ for some $J_e\le K$ (empty parts are discarded). Sorting these parts  from largest to smallest, we obtain $e_{(1)}, \dots, e_{(J_e)}$  with $|e_{(1)}|\ge  \dots \ge | e_{(J_e)}| \ge 1$. In other words, $e_{(k)}$ is the $k$-th largest non-empty part  in $e\cap \calC$. Then we let $p_k=|e_{(k)}|$ be the sizes of these size-ordered parts. The vector  $\bp=(p_1,\dots , p_{J_e})$ belongs to the set of partition vectors 
\[
\calP = \{\bp=(p_1,\dots p_J) ; p_1\ge  \dots \ge p_J \ge 1, \text{ for some } J\ge 1\}.
\]
More generally, for any partition $(e_1,\dots, e_J)$ of a subset of nodes $e\subset V$, we consider its ordering $(e_{(1)},\dots, e_{(J)})$ in size-decreasing order, and let 
\[
\phi(e_1,\dots, e_J)  = (|e_{(1)}|, \dots, |e_{(J)}|) \in \calP.
\]
Note that $\|\phi(e_1,\dots, e_J)\|_1= \sum_k |e_{(k)}|=\sum_k |e_{k}|=|e|$ and we say that the partition vector $\phi(e_1,\dots, e_J)$ has size $|e|$.  
For any partition $\calC$ of the nodes, any subset of nodes $e\subset V$ and any partition vector $\bp\in \calP$, we say that $e$ is \emph{split} by $\calC$ into $\bp$ whenever $\phi(e\cap \calC )= \bp$. In the following, the modularity of a partition $\calC$ will focus on how many hyperedges in $H$ are split into different partition vectors $\bp \in \calP$.

In order to fully understand the links between all the modularities, we introduce a set of nodes subsets as follows.
For any partition $\calC=(C_1,\dots, C_K)$ of the set of nodes $V$ and any partition vector $\bp\in \calP$, with $\|\bp\|_0=J \le K$ and $\|\bp\|_1=s \le n$,  we let  
\[
C_{\bp}= \{ e= \{v_1,\dots, v_s\} \subset V ; \phi(e \cap \calC)= \bp \},
\]
the set of $s$-tuples of nodes that are split into the partition vector $\bp$ by the partition $\calC$. 
With a slight abuse of notation, we can extend the definitions $e_H(\cdot)$ and $\vol_H(\cdot)$ to this set of nodes subsets (while those functions were originally defined on a unique subset of nodes). Thus we let 
\begin{equation*}
e_H(C_{\bp} )= \sum_{e \in \calE} w(e) 1 \{e \in C_{\bp} \}= \sum_{e \in \calE} w(e)1\{\phi(e\cap \calC)= \bp\} ,
\end{equation*}
and 
\begin{align}
\vol_H(C_{\bp} ) 
&:= \sum_{e= \{v_1,\dots, v_s\} \subset V} 1\{\phi(e \cap \calC)=\bp\} \times \prod_{i=1}^s \vol_H(\{v_i\}\cap \calC) \nonumber \\
&= \sum_{\substack {l_1,\dots, l_J \in \{1,\dots,K\} \\ l_j \text{distinct}}} \prod_{j=1}^J \vol_H(C_{l_j})^{p_j} .
\label{eq:volp}
\end{align}
Note that while the quantity $e_H(C_{\bp} )$ is obtained as a direct generalization of the function $e_H$ defined on a unique subset of nodes, the quantity $\vol_H(C_{\bp} )$ is a bit more involved. 
It is a sum-product of volumes over all distinct clusters labels $(l_1,\dots, l_J)$ that induce the same partition $\bp$. These quantities are the generalizations of the former  $\vol_H(C_k)^s$ that appeared when considering only size-$s$ hyperedges  with all nodes belonging to a unique cluster $C_k$.

Let us consider an example to better understand $\vol_H(C_{\bp} )$. We fix a partition $\calC$ of the nodes with $K\ge 3$ parts. Then, the only partition vectors of size $s=2$ are $\bp=(2)$ and $\bp=(1,1)$. Similarly, the only partition vectors of size $s=3$ are $\bp=(3) ; (2,1) ; (1,1,1)$. Now the volumes $\vol_H(C_{\bp}) $ are:
\begin{align*}
\vol_{H}(C_{(2)}) &=  \sum_{k=1}^K  \vol_H(C_k)^2 ; \qquad \vol_{H}(C_{(1,1)}) =  \sum_{k\neq l \in \{1,\dots, K\}} \vol_H(C_k)\vol_H(C_l) ;\\
\vol_{H}(C_{(3)}) &=   \sum_{k=1}^K  \vol_H(C_k)^3 ; \quad 
 \vol_{H}(C_{(2,1)}) =  \sum_{k\neq l \in \{1,\dots, K\}} \vol_H(C_k)^2\vol_H(C_l) ; \\   
 \vol_{H}(C_{(1,1,1)}) &=    \sum_{\substack {k, l,m \in \{1,\dots,K\} \\ k, l, m \text{ distinct}}}  \vol_H(C_k)\vol_H(C_l) \vol_H(C_m) .    
\end{align*}

Note that for any partition $\calC$, any size $s$ and any integer $c\in \{\lfloor s/2\rfloor +1,\dots,s\}$, we have the relation 
\[
\sum_{k=1}^K e_{H}^{s, c}(C_k) = \sum_{\bp\in \calP; \|\bp\|_1=s, p_1=c} e_H(C_{\bp}).
\]
In other words, the  quantity $e_{H}^{s, c}(C_k)$ counts the number of hyperedges of size $s$ having  the majority of their nodes ($c$ of them) in part $C_k$ under partition $\calC$. When summing this over all possible $k$, we get all possible splits of size-$s$ hyperedges into partition vectors $\bp$ such that the majority part has exactly $c$ elements ($p_1=c$).
\\

\cite{Chodrow_21} also introduce a general affinity function $\Omega: \calP \to \R$ that modulates the weight of the contribution of each partition vector $\bp$. They define the symmetric modularity as:
\begin{align*}
\Chsymw(H,\calC) &= \sum_{\bp \in \calP} \Big(\sum_{e \in \calE} w(e) 1\{\phi(e\cap \calC)= \bp\} \log(\Omega(\bp))-\vol_H(C_{\bp})\Omega(\bp)\Big) \\
&= \sum_{\bp \in \calP} \Big(e_H(C_{\bp})\log(\Omega(\bp))-\vol_H(C_{\bp})\Omega(\bp)\Big) .
\end{align*}
A first term in this modularity counts how many (weighted) hyperedges are split by $\calC$ into the different partition vectors $\bp\in \calP$, while a second term is related to the generalized volume $\vol_H(C_{\bp})$. The extra weights $ \log(\Omega(\bp))$ and $ \Omega(\bp)$ might not seem natural at first. In fact, they appear as the result of an approximate maximum likelihood approach in a specific degree-corrected hypergraph stochastic blockmodel (DCSHBM), in the same way as \cite{newm:16}  did in a graph context. As a result, these extra weights $ \log(\Omega(\bp))$ and $ \Omega(\bp)$ modulate the influence of the 2 terms whose differences are summed. Also, (up to the scaling factors $ \log(\Omega(\bp))$ and $ \Omega(\bp)$), while the first term $e_H(C_{\bp})$ in each sum still corresponds to an observed number of specific hyperedges, the second term $\vol_H(C_{\bp})$ is not explicitly its expected value under some probabilistic null model. Nonetheless it is a correction term naturally arising from the consideration of a specific probabilistic model (the DCHSBM).

In practice, the affinity function $\Omega$ is estimated from an initial partition $\calC^{\init}$ through the (weighted) number  of hyperedges split into each partition vector by $\calC^{\init}$, normalized by the volume of this partition vector under $\calC^{\init}$: 
\[ 
\forall \bp \in \calP, \quad \widehat{\Omega}(\bp) = \frac{\sum_{e\in \calE} w(e) 1\{\phi(e\cap \calC^{\init})= \bp\}}{ \vol_H(C_{\bp}^{\init})} = \frac{e_H(C^{\init}_{\bp})}{\vol_H(C_{\bp}^{\init})},
\]
and the modularity becomes  
\begin{equation}
\label{eq:ChodSym_mod}
\Chsym(H,\calC) = 
\sum_{\bp \in \calP} \Big(e_H(C_{\bp}) \log(\widehat{\Omega}(\bp))-\vol_H(C_{\bp}) \widehat{\Omega}(\bp)\Big) .
\end{equation}

The modularity $\Chsym$ represents a compromise between the 2 extremes $\Ku$ and $\Ka$, that goes beyond the one proposed by $\Kawdc$: all hyperedges play a role in this modularity, with weights depending on which partition vector $\bp$ they are split in by partition $\calC$. This is at the cost of estimating many extra affinity parameter $\Omega(\bp)$. 

Currently, the code provided by \cite{Hyper-modularity} does not contain an implementation of an estimation of a general affinity function $\widehat{\Omega}$. Indeed, such a general affinity function requires as many parameters as the total number of possible partitions of a size-$s$ tuple of nodes for any $s\in \{2, \dots, S\}$ with $S=\max_{e \in \calE} |e|$, which is huge. 
Note that their \texttt{demos} directory only contains in the file \texttt{parameterized-affinities.ipynb} an \emph{experimental} version of a very specific parametrized affinity function.

\section{All-or-nothing modularity revisited}
\label{sec:SM_AON}
With the previous notation at stake, we may give a different presentation of $\ChAON$.
The all-or-nothing affinity function $\Omega$ is defined by
\begin{equation}
\label{eq:AON_aff}
\forall \bp \in \calP \text{ such that } \| \bp\|_1= s,  \quad \Omega^{\aon}(\bp) =
\begin{cases}
\omega_{s1} & \text{ if } \|\bp\|_0 =1 ,\\
\omega_{s0} & \text{ else}.
\end{cases}
\end{equation}
Here, the $\ell_1$ norm of an integer vector $\bp=(p_1,\dots, p_J)$ is the sum of its entries $\|\bp\|_1=\sum_j p_j$ and its $\ell_0$ norm is the number of its entries $\|\bp\|_0=J$.
In other words,  a partition vector $\bp$ has AON affinity 
\[
\Omega^{\aon}(\bp)= \sum_{s \ge 2} \Big( \omega_{s1} 1\{\bp=(s)\} + \omega_{s0}1\{\|\bp\|_0\ge 2 ; \|\bp\|_1=s\} \Big).
\]
For each size $s$ of a partition vector, this affinity function is parametrized by only two different values $\omega_{s1}, \omega_{s0}$. These parameters will modulate differently the contributions of hyperedges $e$ depending on their size $s=|e|$ and on whether all their nodes belong to the same cluster (i.e., $\|\phi(e\cap \calC)\|_0=1$) or belong  to at least 2 different clusters  (i.e., $\|\phi(e\cap \calC)\|_0\ge 2$). 
In particular, the volumes~\eqref{eq:volp} computed for partition vectors $\bp$ such that $\|\bp\|_0=1$ write 
\begin{equation*}  
\vol_{H} (C_{(s)}) = \sum_{k=1}^K    \vol_H(C_k)^s .
\end{equation*}
The AON affinity function will in practice be estimated from an initial partition $\calC^{\init}$ through 
 \[ 
 \forall s \ge 2, \quad 
\widehat{\omega}_{s1} = \frac{e_H(C^{\init}_{(s)})}{ \sum_{k=1}^K   \vol_H(C_k^{\init})^s  }  ; \quad 
\widehat{\omega}_{s0} = \frac{\sum_{e\in \calE_s} w(e) 1\{\|\phi(e\cap \calC^{\init})\|_0\ge 2\}}{\sum_{\bp\in \calP ; \|\bp\|_0\ge 2, \|\bp\|_1=s} \vol_{H}(C^{\init}_{\bp})}. 
\]
Finally, inserting that affinity function inside $\Chsymw$ and after some algebra, the resulting modularity becomes (up to  additional constants that do not depend on $\calC$ and are thus neglected)
\begin{align}
\label{eq:ChodAON_mod_v2}
\ChAON (H,\calC) &= -\sum_{s\ge 2} \hat \beta_s \Big( \cut_s(\calC) + \hat \gamma_s \sum_{k=1}^K (\vol_H(C_k))^s \Big)  +c \\
&= \sum_{s\ge 2} \hat \beta_s \Big(e_H(C_{(d)}) - \hat \gamma_s \sum_{k=1}^K (\vol_H(C_k))^s 
\Big) +c  \nonumber \\
&= \sum_{k=1}^K \sum_{s\ge 2} \hat \beta_s \Big(  \sum_{C_k' \subset C_k; |C_k'|=s}e_H(C_k') - \hat \gamma_s (\vol_H(C_k))^s \Big) +c , \nonumber 
\end{align}
where $\hat \beta_s=\log \hat\omega_{s1} -\log \hat \omega_{s0}$ and $\hat \gamma_s= \hat \beta_s^{-1} (\hat\omega_{s1} - \hat \omega_{s0})$, 
the cut terms $\cut_s(\calC)$ count the (weighted) number of hyperedges of size $s$ that contain nodes in two or more distinct clusters; namely, $\cut_s(\calC)= |\calE_s| -e_H(C_{(s)})$.
The first line in Equation~\eqref{eq:ChodAON_mod_v2} is the expression of $\ChAON$ given in \cite{Chodrow_21} while the following lines correspond to our rewriting, showing the similarities with the other previously defined modularities. 
While in general we may expect that $\hat \omega_{s1} >\hat \omega_{s0}$ so that both $\hat \beta_s, \hat \gamma_s >0$, we then recover in this expression a sum of difference terms between a count of specific hyperedges (those entirely included in a cluster) and a correcting volume term.

\section{Linking parameters in HSBM and DCHSBM-like}
\label{sec:SM_HSBM_DC}
If we consider a DCHSBM-like model generating multiset hypergraphs with equal cluster proportions, then we get using twice Baye's rule,  
\begin{align*}
p_s &= \frac{\pr(e=\{v_1,\dots,v_s\} \in \calE_s ; \exists 1\le k\le K,  \phi(e\cap \calC^{\text{true}}) \subset C_k^{\text{true}} ) }{|\calE_s| /n^{s}} \\
&= \frac{\alpha_s K (K/n)^s}{|\calE_s| /n^{s}} = \frac{\alpha_s K^{s+1}}{|\calE_s| } .
\end{align*}
On the other way round, if we consider a HSBM generating simple hypergraphs with cluster proportions $\pi_k$, then we get using twice Baye's rule,
\begin{align*}
\alpha_s = \frac{p_s [\alpha_s \sum_k \pi_k^s + \beta_s (1-\sum_k \pi_k^s)]}{\sum_k \pi_k^s}, 
\end{align*}
 and in the particular case of equal cluster proportions, namely $\pi_k=1/K$, then this leads to
 \[
\alpha_s = \frac{p_s \beta_s(K^{s-1}-1)}{1-p_s}.
\] 

\section{Values for ground truth modularity}
Table~\ref{tab:GT_mod}  shows mean and standard deviations values of the modularities at the ground truth clustering for each method (each one focusing on a specific modularity definition, as summarized in Table~\ref{tab:algos}) and each scenario. 
We observe that the \texttt{IRMM} algorithm that maximizes $\Ku$ has a ground truth modularity close to 0. 

\begin{center}
\begin{longtable}{c|cccc}

 \hline
scenario / method& AON-HMLL & CNM & IRMM & LSR \\ 
  \hline
\endfirsthead

\multicolumn{5}{c}%
{{\bfseries \tablename\ \thetable{} -- continued from previous page}} \\
 \hline
scenario / method& AON-HMLL & CNM & IRMM & LSR \\ 
  \hline
\endhead

\hline \multicolumn{5}{|r|}{{Continued on next page}} \\ \hline
\endfoot

\hline \hline
\endlastfoot

\textbf{ScenA-HSBM} &&&&\\
A1: $n=50$ & -1440.28(253.46) & 0.31(0.04) & -0.01(0.02) & 0.23(0.04) \\ 
  A2: $n=100$ & -3281.04(321.45) & 0.34(0.02) & -0.01(0.02) & 0.27(0.02) \\ 
  A3: $n=150$ & -5089.41(349.39) & 0.35(0.02) & 0(0.01) & 0.27(0.02) \\ 
  A4: $n=200$ & -7007.19(431.96) & 0.36(0.02) & 0(0.01) & 0.28(0.02) \\ 
  A5: $n=500$ & -19151.93(657.66) & 0.36(0.01) & 0(0.01) & 0.28(0.01) \\ 
   \hline
\textbf{ScenA-DCHSBM} &&&&\\
A1: $n=50$ & -1507.05(140.8) & 0.38(0.03) & -0.02(0.02) & 0.3(0.03) \\ 
  A2: $n=100$ & -3214.43(259.46) & 0.37(0.02) & 0(0.02) & 0.29(0.02) \\ 
  A3: $n=150$ & -5055.25(307.23) & 0.37(0.02) & -0.01(0.01) & 0.29(0.02) \\ 
  A4: $n=200$ & -6819.27(297.63) & 0.37(0.02) & 0(0.01) & 0.29(0.02) \\ 
  A5: $n=500$ & -18619.63(568.68) & 0.36(0.01) & 0(0.01) & 0.29(0.01) \\ 
  A6: $n=1000$ & -39925.19(782.67) & 0.36(0.01) & 0(0.01) & 0.29(0.01) \\ 
   \hline
\textbf{ScenA-hABCD} &&&&\\
A1: $n=50$ & -177.69(39.81) & 0.27(0.1) & -0.02(0.07) & 0.24(0.09) \\ 
  A2: $n=100$ & -439.39(58.81) & 0.34(0.07) & -0.01(0.04) & 0.3(0.06) \\ 
  A3: $n=150$ & -727.05(63.2) & 0.36(0.04) & -0.01(0.03) & 0.31(0.03) \\ 
  A4: $n=200$ & -1040.2(69.62) & 0.38(0.04) & -0.01(0.03) & 0.33(0.04) \\ 
  A5: $n=500$ & -3453.7(312.77) & 0.42(0.01) & 0(0.02) & 0.36(0.01) \\ 
  A6: $n=1000$ & -8578.86(633.97) & 0.44(0.01) & 0(0.01) & 0.37(0.01) \\ 
   \hline
\textbf{ScenB-DCHSBM} &&&&\\
B1: $n=50$ & -4476.24(286.51) & 0.38(0.02) & -0.01(0.01) & 0.3(0.02) \\ 
  B2: $n=100$ & -9895.18(438.98) & 0.37(0.01) & -0.01(0.01) & 0.29(0.01) \\ 
  B3: $n=150$ & -15436.71(407.27) & 0.37(0.01) & 0(0.01) & 0.29(0.01) \\ 
  B4: $n=200$ & -21146.49(497.66) & 0.36(0.01) & 0(0.01) & 0.29(0.01) \\ 
  B5: $n=500$ & -57396.92(1005.76) & 0.37(0) & 0(0) & 0.29(0) \\ 
  B6: $n=1000$ & -122253.16(1071.73) & 0.36(0) & 0(0) & 0.29(0) \\ 
   \hline
\textbf{ScenC-HSBM} &&&&\\
C1: $n=50$ & -1692.96(341.87) & 0.22(0.04) & -0.01(0.02) & 0.16(0.04) \\ 
  C2: $n=100$ & -3818.35(526.14) & 0.23(0.03) & -0.01(0.01) & 0.16(0.03) \\ 
  C3: $n=150$ & -6006.44(720.38) & 0.24(0.03) & 0(0.01) & 0.17(0.03) \\ 
  C4: $n=200$ & -8411.32(678.46) & 0.24(0.02) & 0(0.01) & 0.17(0.02) \\ 
  C5: $n=500$ & -23301.14(1133.75) & 0.24(0.02) & 0(0.01) & 0.17(0.02) \\ 
   \hline
\textbf{ScenD-DCHSBM} &&&&\\
D1: $n=50$ & -3268.27(172.98) & 0.47(0.03) & -0.01(0.02) & 0.28(0.02) \\ 
  D2: $n=100$ & -7377.67(213.42) & 0.45(0.02) & 0(0.01) & 0.27(0.01) \\ 
  D3: $n=150$ & -11692.36(325.51) & 0.45(0.02) & -0.01(0.01) & 0.27(0.01) \\ 
  D4: $n=200$ & -15946(274.87) & 0.45(0.01) & 0(0.01) & 0.27(0.01) \\ 
  D5: $n=500$ & -43675.15(431.82) & 0.45(0.01) & 0(0.01) & 0.27(0.01) \\ 
  D6: $n=1000$ & -92887.89(1079.8) & 0.45(0.01) & 0(0) & 0.27(0.01) \\ 
   \hline
\textbf{ScenE-DCHSBM} &&&&\\
E1: $n=50$ & -1496.2(102.41) & 0.41(0.03) & -0.01(0.02) & 0.33(0.02) \\ 
  E2: $n=100$ & -3210.91(190.09) & 0.41(0.02) & 0(0.01) & 0.33(0.02) \\ 
  E3: $n=150$ & -5117.84(249) & 0.41(0.02) & -0.01(0.01) & 0.32(0.02) \\ 
  E4: $n=200$ & -6902.04(311.24) & 0.4(0.02) & -0.01(0.01) & 0.32(0.01) \\ 
  E5: $n=500$ & -19069.69(632.72) & 0.4(0.01) & 0(0.01) & 0.32(0.01) \\ 
  E6: $n=1000$ & -40020.25(699.77) & 0.4(0) & 0(0) & 0.32(0) \\ 
   \hline
\textbf{ScenF-DCHSBM} &&&&\\
F1: $n=50$ & -1401.74(122.37) & 0.33(0.03) & -0.02(0.02) & 0.26(0.02) \\ 
  F2: $n=100$ & -3045.16(240.8) & 0.32(0.02) & 0(0.02) & 0.25(0.02) \\ 
  F3: $n=150$ & -4990.03(275.21) & 0.32(0.02) & 0(0.01) & 0.25(0.01) \\ 
  F4: $n=200$ & -6623.92(208.33) & 0.33(0.02) & 0(0.01) & 0.26(0.02) \\ 
  F5: $n=500$ & -18654.39(359.67) & 0.32(0.01) & 0(0.01) & 0.25(0.01) \\ 
  F6: $n=1000$ & -39341.82(941.03) & 0.32(0.01) & 0(0) & 0.25(0.01) \\ 
   \hline
\textbf{ScenZ-hABCD} &&&&\\
Z1: $n=100$ & -529.92(164.11) & 0.28(0.07) & -0.02(0.04) & 0.26(0.08) \\ 
  Z2: $n=150$ & -782.78(209.52) & 0.33(0.05) & -0.02(0.04) & 0.31(0.05) \\ 
  Z3: $n=200$ & -961.22(128.09) & 0.35(0.04) & -0.01(0.02) & 0.34(0.03) \\ 
  Z4: $n=500$ & -2457.49(132.12) & 0.39(0.02) & 0(0.01) & 0.41(0.02) \\ 
  Z5: $n=1000$ & -5027.19(177.84) & 0.41(0.01) & 0(0.01) & 0.43(0.01) \\ 
   \hline

\caption{Mean value (and standard deviation) of ground truth modularities  $Q^{\textrm{true}}= Q(H, \calC^{\textrm{true}})$ for the 25 hypergraphs generated under each method  and scenario.} 
\label{tab:GT_mod} \\
\end{longtable}
\end{center}

\end{document}